\documentclass{aa}  

\usepackage{graphicx}
\usepackage{amsmath}
\usepackage{verbatim}
\usepackage{fancyvrb}
\usepackage{tikz}
\usetikzlibrary{arrows.meta}
\usetikzlibrary{bayesnet}
\usepackage{txfonts}
\usepackage{natbib}
\bibpunct{(}{)}{;}{a}{}{,} 
\usepackage{color}
\usepackage{xcolor}
\usepackage[colorlinks=true, linkcolor=blue, citecolor=blue, urlcolor=blue]{hyperref}
\usepackage{xspace}
\usepackage{listings}
\usepackage{siunitx}
\definecolor{codegreen}{rgb}{0,0.6,0}
\definecolor{codegray}{rgb}{0.5,0.5,0.5}
\definecolor{codepurple}{rgb}{0.58,0,0.82}
\definecolor{backcolour}{rgb}{0.95,0.95,0.92}

\lstdefinestyle{mystyle}{
    backgroundcolor=\color{backcolour},   
    commentstyle=\color{codegreen},
    keywordstyle=\color{magenta},
    numberstyle=\tiny\color{codegray},
    stringstyle=\color{codepurple},
    basicstyle=\ttfamily\footnotesize,
    breakatwhitespace=false,         
    breaklines=true,                 
    captionpos=b,                    
    keepspaces=true,                 
    numbers=left,                    
    numbersep=5pt,                  
    showspaces=false,                
    showstringspaces=false,
    showtabs=false,                  
    tabsize=2
}

\newcommand\gaia{\textit{Gaia}\xspace}
\newcommand\gdr[1]{\gaia~DR#1\xspace}
\newcommand\edr[1]{\gaia~EDR#1\xspace}

\providecommand\gdr[1]{\textit{Gaia}~DR#1\xspace}

\providecommand\secref[1]{Sect.~\ref{#1}}

\providecommand\secrefalt[1]{Section~\ref{#1}}
\providecommand\figref[1]{Fig.~\ref{#1}}

\providecommand\figrefalt[1]{Figure~\ref{#1}}

\providecommand\equref[1]{Eq.~\eqref{#1}}

\defcitealias{2020MNRAS.497.4246B}{B\&E20}
\defcitealias{2022MNRAS.509.6205E}{E\&B22}
\defcitealias{2023A&A...669A..55C}{CG+23}

\makeatletter
\DeclareRobustCommand*{\fieldName}[1]{%
  \begingroup\@fieldName\scantokens{\texttt{\small {#1}}\noexpand}\endgroup}
\begingroup\lccode`\~=`\_\relax
   \lowercase{\endgroup\def\@fieldName{\catcode`\_=\active \let~\_}}
\makeatother

\begin{document}

\title{Forward model of the \gdr{3} selection function}

\author{
      Alfred~Castro-Ginard          \inst{\ref{inst:barcelona},\ref{inst:iccub},\ref{inst:ieec},\ref{inst:leiden}}\relax
\and  Anthony~G.A.~Brown            \inst{\ref{inst:leiden}}\relax
\and  Eduardo~Belotto~Evangelista   \inst{\ref{inst:barcelona},\ref{inst:iccub},\ref{inst:ieec}}\relax
\and  Jordi~Portell                 \inst{\ref{inst:iccub},\ref{inst:ieec}}\relax
\and  Andrew~R.~Casey               \inst{\ref{inst:cca},\ref{inst:monash}}\relax
\and  Ronald~Drimmel                \inst{\ref{inst:torino}}\relax
\and  Elena~Sellentin               \inst{\ref{inst:leiden-maths},\ref{inst:leiden}}\relax
\and  Tristan~Cantat-Gaudin         \inst{\ref{inst:mpia}}\relax
\and  Morgan~Fouesneau              \inst{\ref{inst:mpia}}\relax
\and  Shourya~Khanna                \inst{\ref{inst:torino}}\relax
\and  Evgeny~P.~Kurbatov            \inst{\ref{inst:cambridge}}\relax
\and  Adrian~M.~Price-Whelan        \inst{\ref{inst:cca}}\relax
}

\institute{Departament de Física Quàntica i Astrofísica (FQA), Universitat de Barcelona (UB), Martí i Franquès, 1, 08028 Barcelona, Spain \email{acastro@fqa.ub.edu}\relax \label{inst:barcelona}
\and{Institut de Ciències del Cosmos (ICCUB), Universitat de Barcelona (UB), Martí i Franquès, 1, 08028 Barcelona, Spain \label{inst:iccub}}
\and{Institut d'Estudis Espacials de Catalunya (IEEC), Edifici RDIT, Campus UPC, 08860 Castelldefels (Barcelona), Spain \label{inst:ieec}}
\and{Leiden Observatory, Leiden University, Einsteinweg 55, 2333 CC Leiden, the Netherlands \label{inst:leiden}}
\and{Center for Computational Astrophysics, Flatiron Institute, 162 Fifth Avenue, New York, NY 10010, USA \label{inst:cca}}
\and{School of Physics and Astronomy, Monash University, VIC 3800, Australia \label{inst:monash}}
\and{INAF – Osservatorio Astrofisico di Torino, Strada Osservatorio 20, Pino Torinese 10025, Torino, Italy \label{inst:torino}}
\and{Mathematical Institute, Leiden University, Einsteinweg 55, 2333 CC Leiden, the Netherlands \label{inst:leiden-maths}}
\and{Max-Planck-Institut für Astronomie, Königstuhl 17, 69117 Heidelberg, Germany \label{inst:mpia}}
\and{Institute of Astronomy, University of Cambridge, Madingley Road, Cambridge CB3 0HA, UK \label{inst:cambridge}}
}

\date{Received date /
Accepted date}
  
\abstract{
    The completeness of an astronomical catalogue depends on its selection function, which is a combination of the observing strategy, instrumental characteristics, and intrinsic properties of the sources being observed. This is particularly complex for \gaia, whose non-uniform scanning law and onboard prioritisation of bright sources in dense areas make its selection function vary strongly with celestial position and source brightness.
}
{
    We estimate the \gdr{3} source catalogue selection function across the whole sky and over the full magnitude range, using only data from the \gaia catalogue and knowledge of the nominal scanning law as provided by the \gaia Observation Forecast Tool.
}
{
    We develop a hierarchical Bayesian model that estimates the number of times \gaia scanned each celestial position jointly with the source detection efficiency, and fit it via Markov Chain Monte Carlo to the $1\,806\,196\,447$ sources in \gdr{3} with reported celestial positions, $G$ magnitudes, and astrometric matched transit counts.
}
{
    \gdr{3} is essentially complete for $G \in (5,19.5)$ in the on-sky median. After that, the completeness decreases with magnitude according to the scanning law pattern combined with the imprint of crowded regions. The completeness is also lower at the very bright end of the survey, consistent with the known limitations on \gaia observations of bright stars. We find an unexpected completeness increase beyond $G \sim 21$, which we conclude is an artefact caused by extended sources, as confirmed by cross-matching against Euclid's Q1 catalogue.
}
{
    Our approach builds on previous efforts that use the understanding of the \gaia survey strategy to model the survey selection function. Our model does not rely on a precise knowledge of the \gaia scanning law, making it more robust, while retaining the capability of deriving an all-sky selection function without the need to extrapolate from limited external surveys providing `ground-truth'. Treating the scanning law as part of the inference process, and anchoring it to all \gdr{3} sources of all magnitudes on a fine-grained grid on the sky, removes the need for an explicit crowding correction. The resulting selection function, valid everywhere where \gaia data are available, is released as part of the GaiaUnlimited Python package.
}
\maketitle
%


\section{Introduction}
\label{sec:intro}

The scientific conclusions drawn from an astronomical catalogue come from an inference process carried out on the catalogue data, not directly on sources in the sky. A catalogue will likely never contain every source with the properties of interest, but only those sources that satisfy some combination of observing strategy, detection limits, and quality cuts. The probability that a source with given properties enters the catalogue, the selection function, is therefore not a nuisance to be minimised by discarding data until the sample looks complete, but a quantity to be modelled and included in the inference process \citep{2021AJ....162..142R}.

Selection effects have been a concern in astronomy since \citet{1922MeLuF.100....1M}, who showed that a sample limited in apparent magnitude preferentially includes intrinsically luminous objects, biasing the inferred mean absolute magnitude of the population. Modern surveys face a substantially harder version of the same problem. This is particularly true for the \gaia survey \citep{2016A&A...595A...1G}, whose selection function is determined by a complex, non-uniform scanning law and by an onboard source-prioritisation strategy that favours bright sources in crowded fields \citep{2016A&A...595A...1G}. The boundary between detected and undetected sources varies non-trivially with both celestial position and magnitude, in a way that no analytical correction can capture.

Efforts to estimate this selection function have concentrated on two strategies. The first calibrates \gaia's completeness empirically, comparing the \gdr{3} catalogue \citep{2023A&A...674A...1G} directly against a deeper external survey acting as {\em ground truth} \citep[][hereafter \citetalias{2023A&A...669A..55C}]{2023A&A...669A..55C}. \gaia, however, is the largest all-sky survey of the Milky Way at an angular resolution of $\sim0.2$ arcsec, and no comparable deep all-sky survey exists to calibrate against. The calibration in \citetalias{2023A&A...669A..55C} is anchored to the sky regions and magnitude range where an overlap exists with the Dark Energy Camera Plane Survey \citep{2018ApJS..234...39S, 2022AAS...24031806S}. The second strategy uses only \gaia data, and forward-models the catalogue generation process with a Bayesian model \citep{2020MNRAS.497.4246B,2022MNRAS.509.6205E}, applied respectively to \gdr{2} and \edr{3} \citep{2018A&A...616A...1G,2021A&A...649A...1G}. To keep the model tractable, \citet[][hereafter \citetalias{2020MNRAS.497.4246B}]{2020MNRAS.497.4246B} reconstructed \gaia's scanning law from the data, and held it fixed while the selection process is estimated. Because the reconstructed scanning law is tied to the same detections it is meant to explain, the resulting estimated completeness tends to be overestimated (see \secref{sec:completeness-sky} for further explanation).

In this paper we extend the Bayesian forward modelling approach described in \citetalias{2020MNRAS.497.4246B}, by treating the scanning law as unknown and estimating it jointly with the selection process. This joint inference, anchored to all \gaia sources, results in completeness estimates across the full sky and the full \gaia magnitude range. The inference is done over a fine grid on the sky which implicitly accounts for crowding effects. This work is part of the GaiaUnlimited project, which aims to determine selection functions for different \gaia data products, and make them available to the astronomical community through Python tools\footnote{\url{https://gaia-unlimited.org}}.

This paper is organised as follows. \secrefalt{sec:modelbasics} provides some general background to the selection function model. \secrefalt{sec:data} describes the data used and their role during the inference process. The Bayesian model and fitting method are described in \secref{sec:method}. \secrefalt{sec:results} describes the resulting selection function and compares it to the empirical model of \citetalias{2023A&A...669A..55C} and the forward model of \citet[][hereafter \citetalias{2022MNRAS.509.6205E}]{2022MNRAS.509.6205E} for \gdr{3}. Finally, the conclusions are presented in \secref{sec:conclusions}.

\section{Selection function model preliminaries}
\label{sec:modelbasics}

The basis of our selection function model is the assumption \citepalias[following][]{2020MNRAS.497.4246B} that the detection of a source by \gaia can be modelled as a Bernoulli process. In this case, the number of times $k$ a source has been detected (and observed) out of the $n$ occasions on which it could have been detected follows a Binomial distribution, 
\begin{equation}
    P(k\mid n,\theta) = \binom{n}{k}\,\theta^{k}\,(1-\theta)^{n-k}\,,
    \label{eq:binomial}
\end{equation}
where $\theta$ is the probability that a source is detected for any given scan of one of the \gaia telescopes across the source (the detection efficiency), which is estimated as part of the selection function modelling (see \secref{sec:method}). The probability of detection for a given source depends in principle only on its brightness and the properties of \gaia's on-board source detection and confirmation chain \citep[see][for details]{2015A&A...576A..74D}. In practice, whether the detection of a source results in an observation that is eventually used in the data processing depends on many factors on-board the spacecraft and in the on-ground data processing pipelines. For example, source crowding in the sky will lead to a lower probability of obtaining faint source observations due to \gaia's prioritisation of bright sources when on-board CCD read-out resources are limited, while decisions related to data quality in the on-ground processing may lead to discarding sets of observations over a range of magnitudes and celestial positions. Hence, as in \citetalias{2020MNRAS.497.4246B}, we interpret $\theta$ as the \emph{effective} detection efficiency, which accounts for any data lost during data collection or discarded during the on-ground data processing, for crowding effects in the sky, and for any other effects that implicitly lower the probability that the observation of a source is used in the data processing. This approach vastly simplifies the modelling of the selection function. In particular, it allows us to infer the \gaia selection function solely from the data contained in the \gaia catalogue (which provides $k$) coupled with knowledge of the scanning law (which provides $n$).

We stress that $n$ represents the potential number of detections (and observations) of a source and $k$ the number of observations of the source that have actually been used to determine the source parameters (astrometry, photometry, etc.). If $n$ and $k$ are known, one can estimate $\theta$ and thus the selection function. This is the principle used by \citetalias{2020MNRAS.497.4246B} and \citetalias{2022MNRAS.509.6205E}. The value of $k$ is known for the sources observed by \gaia. The value of $n$ as a function of celestial position can be predicted from the \gaia scanning law, which gives the telescope pointings as a function of time.

\gaia's scanning of the sky may be interrupted due to events on-board the spacecraft for certain periods of time. During these times, observations are not collected and this amounts to lowering $n$ from what it could have been according to the \gaia scanning law (losses in the data transmission from \gaia to Earth also affect $n$, but this is a negligible effect). On the other hand, in the on-ground data processing, certain data points for a source may be discarded because they are deemed unusable (for example, data taken during periods of spacecraft thermal instability or during periods of excessive stray light on the focal plane) and this will lower the value of $k$. In an ideal world the \gaia scanning law would be perfectly known, including the data collection gaps, and then with the knowledge of $k$ for each source it is straightforward to estimate $\theta$, which we assume depends only on celestial position and source magnitude, i.e., $\theta=\theta(\ell,b,G)$. The scanning law as provided by the \gaia Observation Forecast Tool (GOST)\footnote{\url{https://gaia.esac.esa.int/gost/}} deviates from the actual scanning law carried out by \gaia, and we opted to estimate the true value of $n$ as part of the selection function modelling process, using the GOST predictions as prior information\footnote{Updating the GOST tool with the known actual orbit of \gaia and the actual spacecraft attitude, as estimated during the astrometric processing, is in principle possible but this was not done due to lack of available effort.}. Because, to first order, the detection efficiency can be estimated as $\theta\sim k/n$, it is important to distinguish between a lowering of $n$ due to data collection interruptions or a lowering of $k$ due to data processing choices. For a given $k$, if $n$ was lowered due to data-taking interruptions, the estimated effective detection efficiency will be higher. This is correct, as the $k$ detections were obtained for a number of scans lower than potentially possible. If, on the other hand, the value of $k$ is lowered for fixed $n$, the effective detection efficiency will correctly be estimated as lower. The context provided in this section should be kept in mind for the following sections.

\section{Data}
\label{sec:data}

We use \gdr{3} \citep{2023A&A...674A...1G}, which corresponds to the $34$ months of observations from July $25$, $2014$ to May $28$, $2017$. We use for every source its celestial position, its $G$ magnitude, and its number of matched transits used in the astrometric solution, reported as \fieldName{astrometric_matched_transits} in the \gaia archive. The presence of a source in the main \gdr{3} table, \fieldName{gaia_source}, implies the availability of an astrometric solution for the source, hence the relevant number of transits used is \fieldName{astrometric_matched_transits} and not \fieldName{matched_transits}\footnote{\fieldName{matched_transits} is the number of field-of-view transits matched to the source, while \fieldName{astrometric_matched_transits} counts only those that were used in the astrometric solution (AGIS).}. This also implies that the observation time range that determines the selection function is 33 months, starting from August 22, 2014 \citep{2021A&A...649A...2L}. We fit the model described in \secref{sec:method} to every \gdr{3} source with $G \in (1,22)$, and for which all three quantities, i.e., celestial position, $G$, and \fieldName{astrometric_matched_transits}, are reported, $1\,806\,196\,447$ sources in total. 

As mentioned above, the GOST tool does not account for interruptions to nominal observations, so it over-predicts the number of scans that are actually usable. Therefore, we use it only as a prior on the true scan count, rather than as the number of scans itself (see \secref{sec:modelbasics} above and \secref{sec:latentn}).

The difference between the predicted and the usable number of scans is documented in \gaia's data-taking gaps tables\footnote{\url{https://www.cosmos.esa.int/web/gaia/dr3-data-gaps}}. These represent the known intervals during which data were not recorded or were discarded in the astrometric or photometric (and spectroscopic, although not used in this work) processing, due to events such as payload decontamination, telescope refocussing, and other anomalies. These gap tables give the retention fraction we adopt in \secref{sec:latentn}. The source density-dependent data losses from the onboard source prioritisation strategy in crowded regions, are recovered implicitly by our model.

\section{Method}
\label{sec:method}

We estimate the \gdr{3} survey selection function, meaning the probability that a source present on the sky enters the catalogue, by forward-modelling the catalogue generation process following the recipe described in \citetalias{2020MNRAS.497.4246B}. As outlined above, \gaia scans each celestial position a number of times $n$ which results in a number of detections $k$ which correspond here to the number of transits matched to a source and used in the astrometric solution (\fieldName{astrometric_matched_transits}). The number of matched transits then follows the distribution given by \equref{eq:binomial}.

A source is present in \gdr{3} if it satisfies the following conditions: it has at least a two-parameter astrometric solution (celestial position) and a $G$ magnitude  and at least five astrometric matched transits are available. Hence, the catalogue is a left-truncated realisation of \equref{eq:binomial}. The requirements for five- or six-parameter solutions are even more restrictive \citep{2021A&A...649A...2L}, but in all cases at least five matched transits must be available. This selection criterion modifies \equref{eq:binomial} to
\begin{equation}
    P(k\mid n,\theta) =
    \begin{cases}
        \dfrac{1}{I_{\theta}(5,\,n-4)}\,\dbinom{n}{k}\,\theta^{k}\,(1-\theta)^{n-k} & k \ge 5\\[2mm]
        0 & \text{otherwise}
    \end{cases}\,,
    \label{eq:truncated}
\end{equation}
where the normalisation is the regularised incomplete Beta function
\begin{equation}
    I_{\theta}(a,b) = \frac{1}{\mathrm{B}(a,b)}\int_{0}^{\theta} x^{a-1}(1-x)^{b-1}\,\mathrm{d}x\,,
    \label{eq:ibeta}
\end{equation}
which accounts for the left-truncation of the binomial distribution at $k = 5$. The normalisation term $\mathrm{B}(a,b)$ is the Beta function, i.e., the same integral evaluated from $0$ to $1$.

We fit this model to every \gdr{3} source with a reported celestial position, $G$ magnitude, and \fieldName{astrometric_matched_transits} value. These sources are binned on the sky with a HEALPix tessellation at level $7$ \citep[$196\,608$ equal-area pixels,][]{2005ApJ...622..759G}, and in magnitude in $0.2$\,mag wide bins over $G \in (1,22)$.

\subsection{Model for the detection efficiency}
\label{sec:effmodel}

Stars at a fixed magnitude may have varying detection efficiencies depending, for instance, on their celestial position or colour. Following \citetalias{2020MNRAS.497.4246B}, we model the distribution of $\theta$ using a Beta distribution with magnitude-dependent parameters,
\begin{equation}
    P(\theta\mid A,B) = \frac{\theta^{A-1}(1-\theta)^{B-1}}{\mathrm{B}(A,B)},
    \label{eq:beta}
\end{equation}
with $A(G)$ and $B(G)$ estimated per HEALPix and magnitude bin. Combining \equref{eq:truncated} and \equref{eq:beta} and marginalizing over $\theta$, results in the following likelihood of $k$ for a given $(n,k)$ pair:
\begin{equation}
    P(k\mid n,A,B) =
    \begin{cases}
        \dbinom{n}{k}\,\dfrac{J(n,k,A,B)}{\mathrm{B}(A,B)} & k \ge 5\\[3mm]
        0 & \text{otherwise}
    \end{cases}\,,
    \label{eq:pnk}
\end{equation}
where $J(n,k,A,B)$ is the integral\footnote{The integral given here corrects a typographical error in the numerator of Eq.~(13) of \citetalias{2020MNRAS.497.4246B}.} 
\begin{equation}
    J(n,k,A,B) = \int_{0}^{1}\frac{\theta^{A+k-1}(1-\theta)^{B+n-k-1}}{I_{\theta}(5,\,n-4)}\,\mathrm{d}\theta\,.
    \label{eq:J}
\end{equation}
Within a HEALPix, the sources that share the same $(n,k)$ are indistinguishable, so we group them and describe the log-likelihood as
\begin{equation}
    \log\mathcal{L} = \sum_{i} c_{i}\,\log P(k_{i}\,|\,n_{i},A,B),
    \label{eq:loglik}
\end{equation}
where $c_i$ counts the sources with $n_i$ scans and $k_i$ astrometric matched transits.

\subsection{Effective number of scans}
\label{sec:latentn}

A reasonably accurate prediction of the \gaia scanning law for an individual sky coordinate is provided by GOST. The number of scans depends on the celestial position. Querying GOST at each source position (for 1.8 billion sources) is computationally expensive, so we opt to determine the selection function on a HEALPix grid at level 7 (NSIDE of 128). At this resolution each pixel spans $0.2$ square degrees and, depending on the celestial position of the pixel, the number of scans can vary significantly over its area. We therefore require an \emph{effective} number of scans for the pixel, rather than the value returned for its central coordinate alone. We estimate the effective scan count $N_{\rm eff}$ as a latent variable per pixel, shared across all magnitude bins, using GOST and the record of data-taking gaps as prior knowledge, as described below.

In our model, we use a discrete uniform prior on $N_{\rm eff}$ centred on the value provided by GOST, $n_{\rm GOST}$, with an empirically determined half-width of $50$ scans, 
\begin{equation}
    N_{\rm eff} \sim \mathcal{U}\{n_{\rm GOST}-50,\;n_{\rm GOST}+50\}.
    \label{eq:nprior}
\end{equation}
For several pixels spread across the sky, we compared $n_{\rm GOST}$ at the centre of the HEALPix level-$7$ pixel with the GOST prediction at each HEALPix level-$12$ coordinate contained in that pixel. We found a maximum difference of $41$ scans, so we adopt a half-width of $50$ to keep the prior wide and uninformative.

Additionally, we know the epochs at which \gaia was not observing, or at which the data were not used in the astrometric and photometric solutions, due to different reasons. These data-taking gaps remove a fraction of the scans, which we encode with a retention fraction $f_{\rm ret}$ giving a usable scan count 
\begin{equation}
    N_{\rm usable} = \lfloor\,f_{\rm ret}\,N_{\rm eff} + 0.5\,\rfloor,
    \label{eq:nusable}
\end{equation}
i.e., $f_{\rm ret}\,N_{\rm eff}$ rounded to the nearest integer. We restrict it to $N_{\rm usable} \geq \max_i\,k_i$, so that no source has more matched transits than scans. When calculating $f_\mathrm{ret}$ we assume all data gaps in the tables provided with \gdr{3} correspond to on-board interruptions of data collection (i.e.\ $n$ is lowered, see \secref{sec:modelbasics}). Given that the GOST predictions are only used as a prior on $N_\mathrm{eff}$, this simplification should not cause significant errors in the inference of the selection function.

The inclusion of this latent variable to account for \gaia's number of scans removes the inconsistency that a small fraction of sources has more matched transits than scans predicted by GOST \citepalias[][their Appendix~B]{2020MNRAS.497.4246B}, and includes these sources as part of the inference process. Additionally, sharing $N_{\rm eff}$ across magnitude bins ties every bin to the same scan count, so the count within a pixel is tightly constrained by all $G$ bins jointly. Because we recover the effective number of scans, rather than a value reconstructed from the detected sources, the reduced detection efficiency of faint sources in crowded fields is reproduced by the fitted efficiency $\theta$ itself. Therefore, our model needs no explicit crowding term. \citetalias{2020MNRAS.497.4246B} took a similar approach to accounting for crowding by inferring detection efficiencies averaged over sky areas covering different source density intervals. Basically, we take their approach further by solving for the detection efficiencies for each HEALPixel on the sky, thus implicitly accounting for crowding effects at the $\sim0.2$~deg$^2$ level.

\begin{figure}
    \centering
    \includegraphics[width=\hsize]{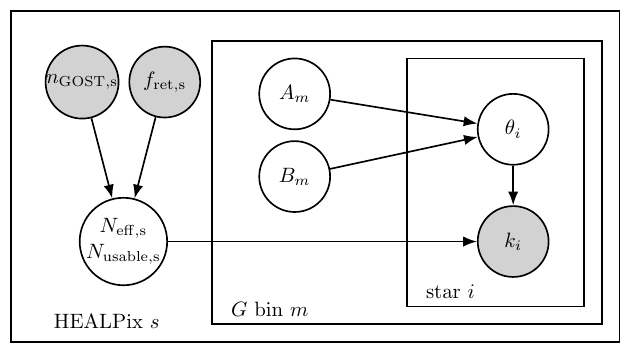}
    \caption{Directed Acyclic Graph (DAG) of the Bayesian hierarchical model for the \gdr{3} source catalogue selection function.
    Shaded nodes correspond to the data used to fit the model: i) $n_{\rm GOST}$,
    ii) $f_{\rm ret}$, and iii) $k$ or \fieldName{astrometric_matched_transits}. Open
    nodes are inferred. The latent number of scans $N_{\rm eff}$ sits outside the
    magnitude-bin plate, so it is shared across all magnitude bins within a pixel. The same node also shows $N_{\rm usable,s}$, its retention-corrected count (\equref{eq:nusable}).}
    \label{fig:model}
\end{figure}

Figure~\ref{fig:model} summarises the resulting model structure. Each plate (square) represents a level of the model hierarchy: i) the HEALPix pixel $s$ which sets the latent scan count $N_{\mathrm{eff},s}$, ii) the magnitude bins $m$ within it, and iii) the individual stars $i$, whose detection efficiency $\theta_i$ is drawn from the Beta distribution parametrised by $(A_{m,s},B_{m,s})$.

\subsection{Inference through MCMC sampling}
\label{sec:inference}

We sample in $(\log_{10}\,A_{m,s},\,\log_{10}\,B_{m,s})$ per magnitude bin $m$ and for each pixel $s$, each with a uniform prior over $[-1,4]$ \citepalias[following][]{2020MNRAS.497.4246B}, and marginalise over the latent scan count. We denote the set of Beta parameters of all magnitude bins in pixel $s$ by 
\begin{equation}
    \boldsymbol{\phi}_s \equiv \big(A_{1,s},B_{1,s},\,\ldots,\,A_{n_{\rm bin},s},B_{n_{\rm bin},s}\big).
    \label{eq:phi}
\end{equation} 
The posterior for pixel $s$ is 
\begin{align}
    p\big(\boldsymbol{\phi}_s\mid\mathrm{data}\big) \propto
    &\prod_{m} p\big(A_{m,s},B_{m,s}\big) \nonumber\\
    &\quad \times \sum_{n}
    \Bigg[\,p\big(N_{\mathrm{eff},s}{=}n\big) \nonumber\\
    &\qquad \times\prod_{m}\prod_{i} P\big(k_{m,i}\mid N_{\rm usable}(n),A_{m,s},B_{m,s}\big)^{c_{m,i}}\Bigg],
    \label{eq:posterior}
\end{align}
where $p(A_{m,s},B_{m,s})$ is the prior on the Beta parameters, $p(N_{\mathrm{eff},s}{=}n)$ is the discrete uniform prior of \equref{eq:nprior}, $m$ runs over the magnitude bins of the pixel, $i$ over the distinct values of $k$ in each bin, $n$ runs over the support of the prior in \equref{eq:nprior}, and $N_{\rm usable}(n)$ is given by \equref{eq:nusable}. Because the latent $N_{\rm eff,s}$ is shared by all magnitude bins, the sum over $n$ cannot be factorised per bin, and the fit is joint across all $m$ bins. In practice, each term in the sum is a product of many small probabilities that would underflow numerically, so we evaluate the sum from the logarithms of its terms using the \mbox{log-sum-exp} operation. The posterior over $\boldsymbol{\phi}_s$ (Eq.~\ref{eq:posterior}), of dimension $2\,n_{\rm bin}$, is sampled with the affine-invariant ensemble sampler \texttt{emcee} \citep{2013PASP..125..306F}, using $16\,n_{\rm bin}$ walkers per pixel initialised from a rough per-bin estimate of the efficiency. We run each chain for $2\,000$ steps, discarding the first $600$ as burn-in and thinning the remainder by a factor of two. Given our priors and sample sizes, the retained chain always exceeds $50$ times the autocorrelation time estimated across all parameters, which is the recommended threshold for convergence.

Although $N_{\mathrm{eff},s}$ is marginalised over in Eq.~\ref{eq:posterior} and is therefore not itself a sampled parameter, we recover its posterior distribution in a post-processing step. For a subsample of posterior draws of $\boldsymbol{\phi}$, we compute the same per-$n$ likelihood terms that enter Eq.~\ref{eq:posterior} and normalise them into a probability distribution over $n$. This gives the full data-driven posterior $p(N_{\mathrm{eff},s}\mid\mathrm{data})$ for every pixel. 

Evaluating $J$ (Eq.~\ref{eq:J}) by quadrature at every sampler step is prohibitive over $\sim 10^9$ sources. Therefore, for each pixel and each relevant $(n,k)$ pair, we precompute $\log J$ on a grid in $(\log_{10}A,\,\log_{10}B)$ using Gauss--Legendre quadrature, and interpolate it during sampling with a bivariate spline. The binomial coefficients, $\log\left[\binom{n}{k}\right]$, are additionally cached per pixel to further increase computation speed. 

\subsection{Code-to-data implementation in SPACIOUS}

The inference is embarrassingly parallel across pixels, and we run it on two different systems. On the one hand, we use the computational resources provided by SPACIOUS (Science PlAtform Cloud Infrastructure for Outsize Usage Scenarios). \mbox{SPACIOUS}\footnote{\url{https://spacious.ub.edu/}} is a Horizon Europe project aiming to provide a cloud-native infrastructure for large-scale astrophysical data analysis. We deploy the pipeline with Apache Spark, dispatching every pixel through \texttt{applyInPandas} keyed on the HEALPix index. On the other hand, we use a traditional Slurm-managed \citep{10.1007/978-3-031-43943-8_1} HPC cluster, where each HEALPix is computed in parallel through a Python \texttt{multiprocessing.Pool} that distributes the pixels across the cores of a single node. We use both deployments for different reasons: i) the Slurm cluster extends the number of available CPUs, while ii) the Spark deployment tests the code-to-data paradigm in which the computation is shipped to the cluster hosting the data rather than the data being downloaded for local compute.

We compare the performance of the two deployments on a set of $60$ HEALPix-7 pixels representing five regions of the sky: bulge, inner disc, anticentre, mid-latitude $b$ (representing the thick disc transition), and Galactic poles. These regions represent a wide range of per-pixel computational cost, which is driven by stellar density. On the Slurm cluster, the fastest pixel completes in $1.32\,\mathrm{hr}$ and the slowest in $13.74\,\mathrm{hr}$, with a mean of $5.05\,\mathrm{hr}$ per pixel. On SPACIOUS, the fastest pixel completes in $0.56\,\mathrm{hr}$ and the slowest in $5.79\,\mathrm{hr}$, with a mean of $2.32\,\mathrm{hr}$ per pixel. The ratio between the two is close to constant at $2.26$ across all pixels in all five regions, which is a signature of a per-core hardware speed difference rather than memory pressure or scheduling. We should note that both sets of timings are compute-only, and the data processing step, i.e., querying, binning, and, for the Slurm cluster, downloading and uploading the \textit{Gaia} data beforehand, is excluded for both deployments. \mbox{SPACIOUS} reaches the same starting point directly from the raw \textit{Gaia} data automatically and within the run, while the Slurm pipeline requires separate manual extraction and upload steps. For future astronomical surveys, however, the data volume will be of the order of Tb or even Pb. For those, the code-to-data paradigm represented by SPACIOUS will not only continue to offer a performance gain but, more importantly, will enable the computation which is intractable in traditional data-to-code approaches. We additionally quantified the cost of this SPACIOUS run by deploying it on the Google Cloud service, where the same computation costs on average €$0.50$ per HEALPix (as of September 2026).

\section{Results}
\label{sec:results}

In this section, we present the resulting \gdr{3} selection function, obtained by combining the estimated parameters in our Bayesian model ($A, B$; see \secref{sec:effmodel}) with the recovered effective scanning law $N_{\rm eff}$ (see \secref{sec:latentn}). For every HEALPix $s$ and magnitude bin $m$, we define \gaia's selection probability $\mathcal{S}_{\textit{Gaia}}\,(s,m)$ as the probability that a source is detected in at least the five matched transits required to be included in the astrometric solution. We obtain it by marginalising over the fitted Beta distribution of $\theta$ \citepalias[Eq.~\ref{eq:beta}, cf.][their equation 6]{2020MNRAS.497.4246B}, and summing the resulting Beta-Binomial probability over all matched transits at or above the five-transit threshold,
\begin{equation}
    \mathcal{S}_{\textit{Gaia}}(s,m) = 1 - \sum_{j=0}^{4}\binom{N_{\mathrm{eff},s}}{j}\,\frac{\mathrm{B}\!\left(A_{m,s}+j,\;B_{m,s}+N_{\mathrm{eff},s}-j\right)}{\mathrm{B}(A_{m,s},B_{m,s})},
    \label{eq:selprob}
\end{equation}
where $N_{\mathrm{eff},s}$, $A_{m,s}$, and $B_{m,s}$ are the outcomes of the inference process.

\subsection{Effective scanning law}
\label{sec:effective-scanning-law}

\begin{figure}
    \centering
    \includegraphics[width=\columnwidth]{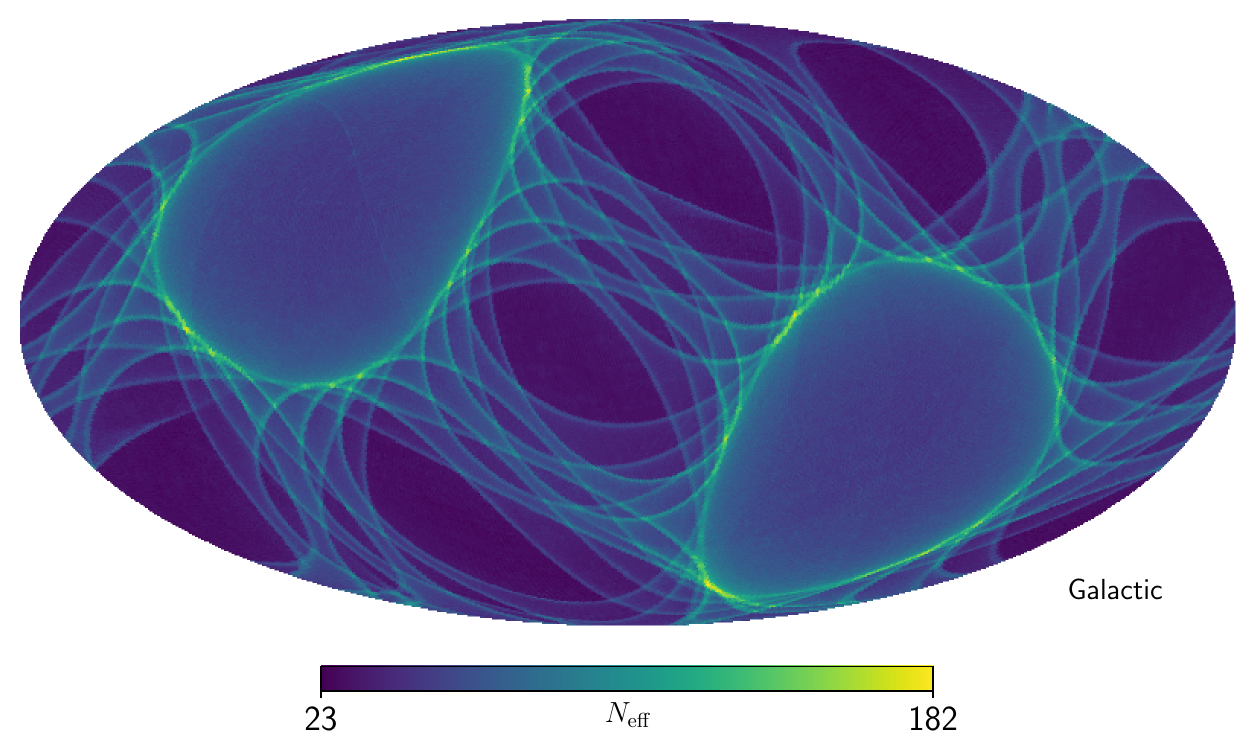}
    \caption{Sky map in Galactic coordinates of the posterior median $N_{\rm eff}$, the effective number of scans per HEALPix at level $7$ resolution. The code to reproduce this plot is in Listing~\ref{lst:gaiaunlimited-scanning-law}.}
    \label{fig:scanning-law}
\end{figure}

As discussed in \secref{sec:latentn}, we recover an effective, data-driven scanning law rather than relying on the GOST prediction alone. This avoids discarding sources in the inference process, and removes biases introduced by a finite spatial resolution. In Fig.~\ref{fig:scanning-law}, we show the effective scanning law for a resolution of HEALPix at level $7$ (corresponding to a pixel area of $0.2$\,deg$^2$), which shows the characteristic \gaia scanning pattern, as expected. The posterior on $N_{\rm eff}$ is tightly constrained across most of the sky, where the posterior width (difference between the $84^{\rm th}$ and $16^{\rm th}$ percentiles) is zero for $\sim 90\%$ of the pixels, and one for $\sim 10\%$ of the pixels. $N_{\rm eff}$'s posterior width is $2$ for $61$ pixels, and greater than $2$ for $10$ pixels with the least constrained having a spread of $8$ scans. This effective scanning law, together with the summary statistics of its posterior and $N_{\rm usable}$, is released alongside the selection function.

\begin{figure}[!hb]
    \centering
    \includegraphics[width=\columnwidth]{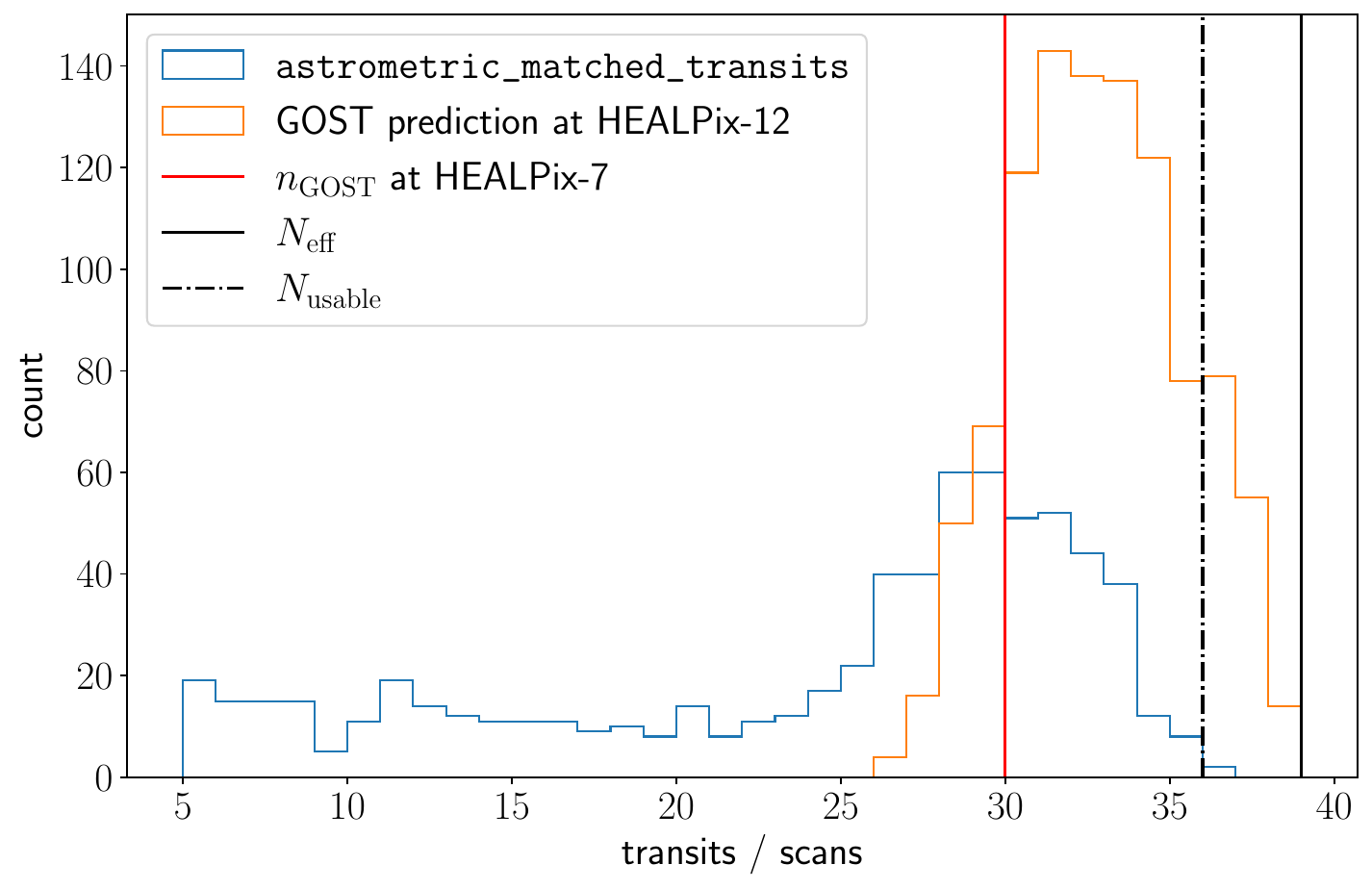}
    \caption{Histogram of \gdr{3} \fieldName{astrometric_matched_transits} (blue) for a pixel centred at $\ell = \ang{153.63}$, $b = \ang{20.10}$, together with the predicted GOST scans at every child HEALPix-12 (orange). The vertical lines show the number of scans estimated by GOST (red line), and our model $N_{\rm eff}$ (black, solid line) and $N_{\rm usable}$ (black, dashed line). The usable fraction of scans ($f_{\rm ret}$) for this pixel is $93\%$.}
    \label{fig:transits-scans}
\end{figure}

\begin{figure*}
    \centering
    \includegraphics[width=\textwidth]{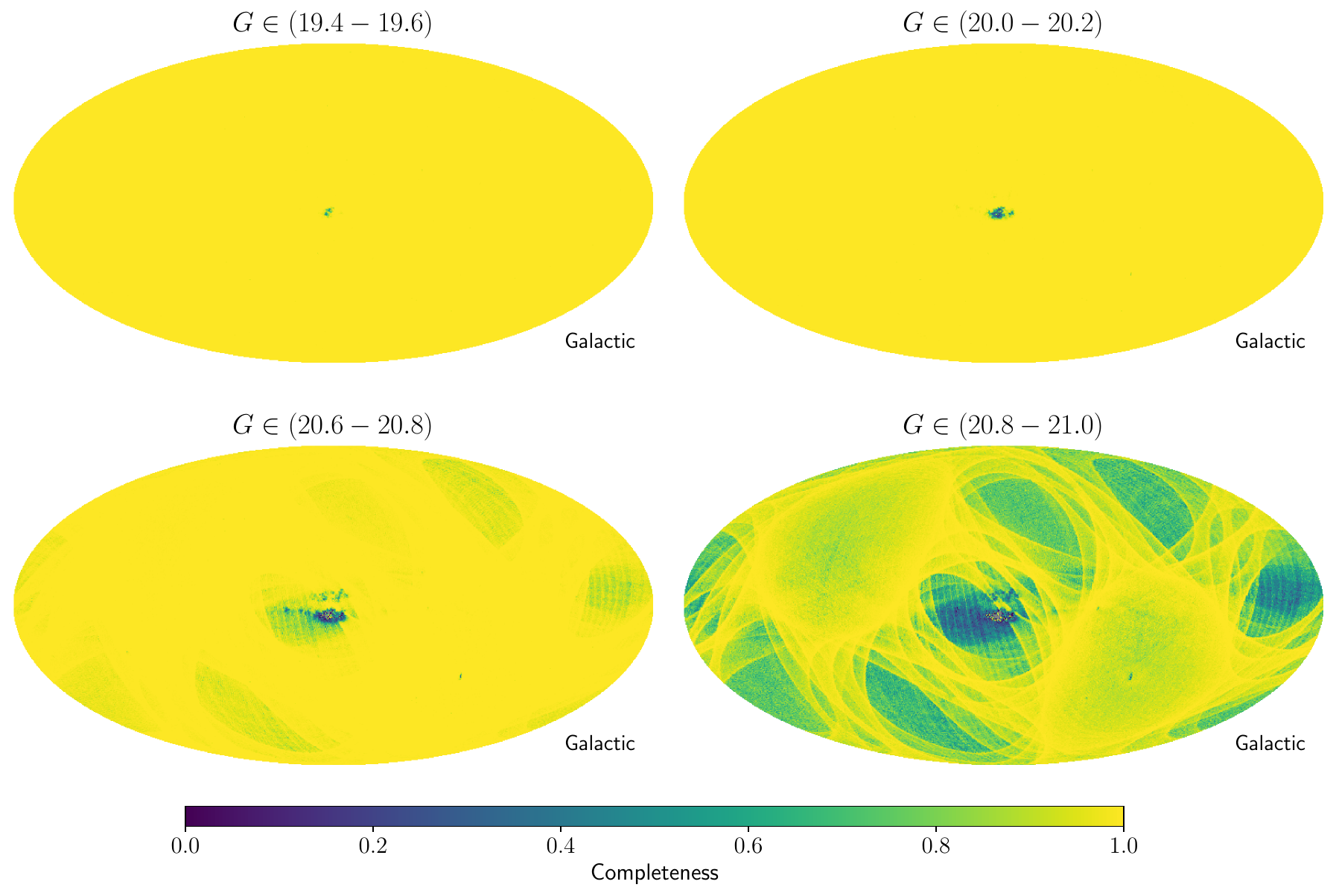}
    \caption{All-sky maps in Galactic coordinates of the selection function at four different magnitude bins towards \gaia's faint end. From top-left to bottom-right: i) $G\in(19.4,19.6)$, ii) $G\in(20.0,20.2)$, iii) $G\in(20.6,20.8)$, and iv) $G\in(20.8,21.0)$. Completeness progressively drops with magnitude, affecting crowded regions first and regions with fewer scans second. The code to reproduce this plot is in Listing~\ref{lst:magnitude-slices}.}
    \label{fig:magnitude-slices}
\end{figure*}

\begin{figure*}
    \centering
    \includegraphics[width=\textwidth]{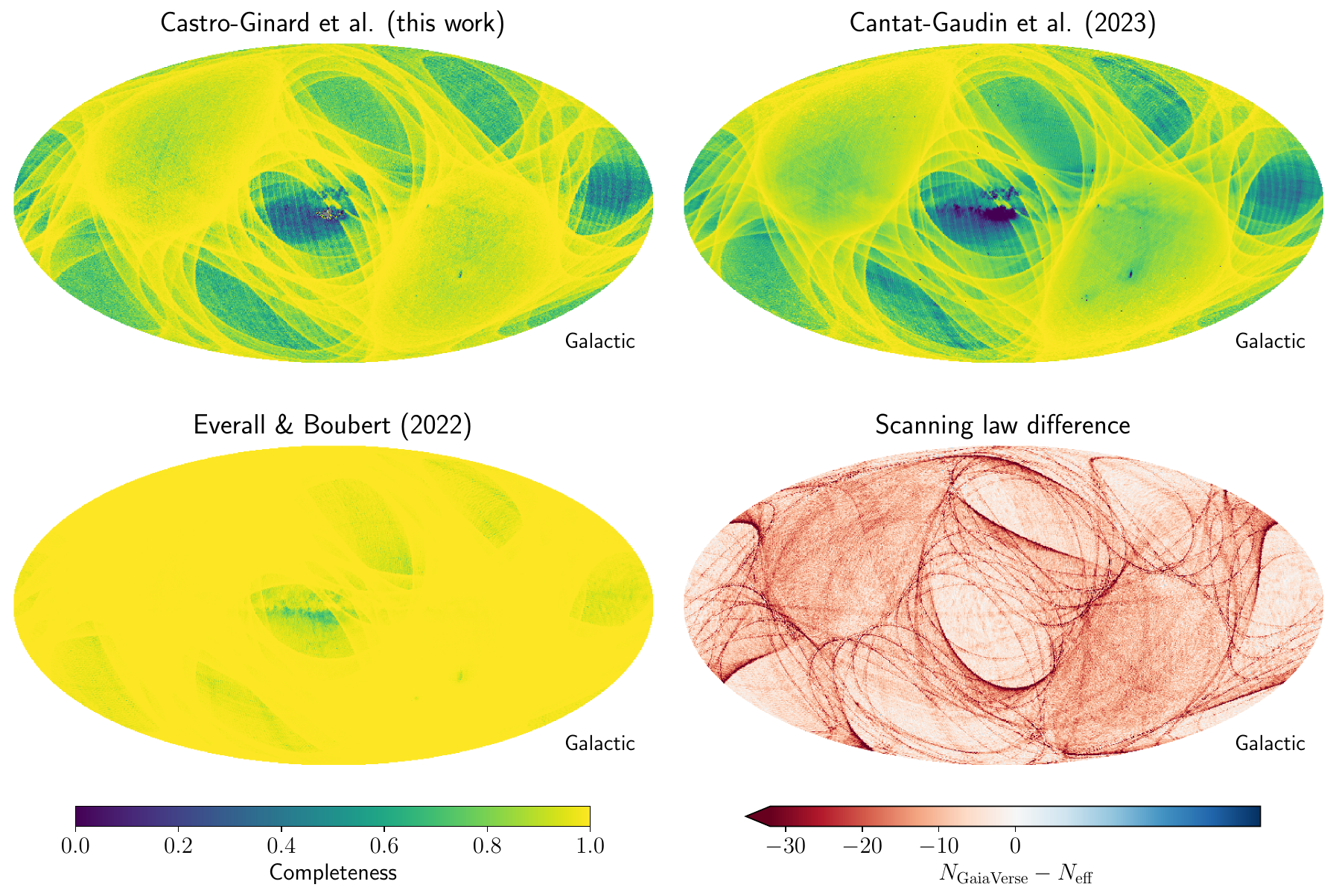}
    \caption{All-sky maps in Galactic coordinates of three different estimations of the \gdr{3} selection function at $G\in(20.8,21)$: i) our forward model, ii) \citetalias{2023A&A...669A..55C} empirical model, and iii) \citetalias{2022MNRAS.509.6205E} forward model. The bottom-right panel shows the differences between our estimation of $N_{\rm eff}$ and the scanning law used by \citetalias{2022MNRAS.509.6205E}.}
    \label{fig:all-comparison}
\end{figure*}

We illustrate in Fig.~\ref{fig:transits-scans} why this effective scanning law is needed by showing the distribution of \fieldName{astrometric_matched_transits} for a level 7 HEALPix at $\ell = \ang{153.63}$ and $b = \ang{20.10}$, together with the different estimations of the number of scans. At HEALPix-7, GOST (red line) predicts a number of scans close to the mode of the observed distribution of matched transits (blue histogram). This would discard the genuine detections above $k = 30$ and overestimate the selection probability if used as a scan count directly. Instead, our estimation of $N_{\rm eff}$ (black, solid line), sits at the maximum of the GOST predictions at HEALPix-12 (orange histogram), and correctly accounts for the resolution element placing the number of scans so that $N_{\rm eff} > k$. Furthermore, only $93\%$ of the scans in this region were used in the astrometric/photometric processing, and not accounting for this would underestimate the selection probability for these sources.

\subsection{Completeness on sky}
\label{sec:completeness-sky}

The main result of this work is the \gdr{3} selection function as a joint function of celestial position and magnitude, and released within the {\tt GaiaUnlimited} Python package (see \secref{sec:gaiaunlimited}). Figure~\ref{fig:magnitude-slices} shows in different all-sky maps how the selection function varies with $G$. At $G \sim 19.5$ (top-left panel), the \gdr{3} catalogue is essentially complete, except for the Galactic centre region where the source selection probability drops to $\sim 50\%$ due to crowding effects. For fainter magnitudes ($20<G<20.8$, top-right and bottom-left panels), the incompleteness due to crowding has both widened and dropped further, reaching $0\%$ at the Galactic centre. In addition, we see how less scanned regions correspond to less complete regions in the sky. When $G$ reaches $21$\,mag, \gdr{3} is largely incomplete, with the scanning law pattern becoming visible and the Galactic centre reaching its lowest completeness (including pixels with no sources).

In Fig.~\ref{fig:all-comparison}, we compare our selection function at $G\in(20.8,21)$ with two all-sky estimates available for the \gdr{3} source catalogue, i.e., the empirical model of \citetalias{2023A&A...669A..55C}, and the forward model of \citetalias{2022MNRAS.509.6205E} who used the \gdr{2} detection efficiencies \citepalias{2020MNRAS.497.4246B} in combination with the \edr{3} scanning law. Our map and that of \citetalias{2023A&A...669A..55C} are very similar, with our selection probability slightly higher (yellower) across the sky. We discuss the origin of this offset below, zooming in on the Galactic centre to highlight the differences. The selection function of \citetalias{2022MNRAS.509.6205E} is systematically higher than both, and very close to $100\%$ across the sky except for the Galactic centre where it falls to $\sim 60\%$.

Both forward models describe the selection in terms of the number of scans and the detection efficiency, but obtain them differently. \citetalias{2022MNRAS.509.6205E} compute the number of scans, $N_{\rm GaiaVerse}$, directly from the nominal \edr{3} scanning law (equivalent to GOST), and the bottom-right panel of Fig.~\ref{fig:all-comparison} shows that it is systematically lower than our effective number of scans $N_{\rm eff}$, particularly in regions where most scans have been lost due to \gaia's data-taking gaps. This is the same effect that we observe in Fig.~\ref{fig:transits-scans} and discuss in Sect.~\ref{sec:effective-scanning-law}. Fewer scans alone would lower the selection probability for a fixed $\theta$, so the higher completeness of \citetalias{2022MNRAS.509.6205E} must come from a combination of fewer scans and higher detection efficiency. This efficiency is not estimated from \edr{3} data, but taken from the \gdr{2} fit of \citetalias{2020MNRAS.497.4246B}, where the parameters $(A,B)$ of the Beta distribution of $\theta$ (Eq.~\ref{eq:beta}) are a function of $G$ only, not of sky position. Crowding was accounted for using a set of fits of $(A,B)$ to sky areas covering 10 ranges in source density, where for each area $A$ and $B$ are only a function of $G$. Every sky position with the same magnitude and in the same interval of source density thus shares the same efficiency. In our model, $(A_{m,s},B_{m,s})$ are estimated independently for every HEALPix and magnitude bin, so the efficiency follows the local scanning and crowding conditions. In addition, \citetalias{2020MNRAS.497.4246B} fitted their efficiency against a scanning law reconstructed from the \gaia detections themselves. A scan count tied to the detections is pulled towards the number of matched transits $k$, and since $\theta\sim k/n$ (\secref{sec:modelbasics}), the fitted efficiency is inflated. Applying this inflated efficiency to the \edr{3} scan counts, which are also underestimated, reproduces the same combination of low $n$ and high $\theta$ in their \edr{3} selection function. We attribute the higher completeness of \citetalias{2022MNRAS.509.6205E} to this inflated efficiency in combination with the fewer scans discussed above, whose effect we expect to be largest for faint sources in crowded fields where the number of matched transits differs the most from the number of scans. 

\begin{figure}
    \centering
    \includegraphics[width=\columnwidth]{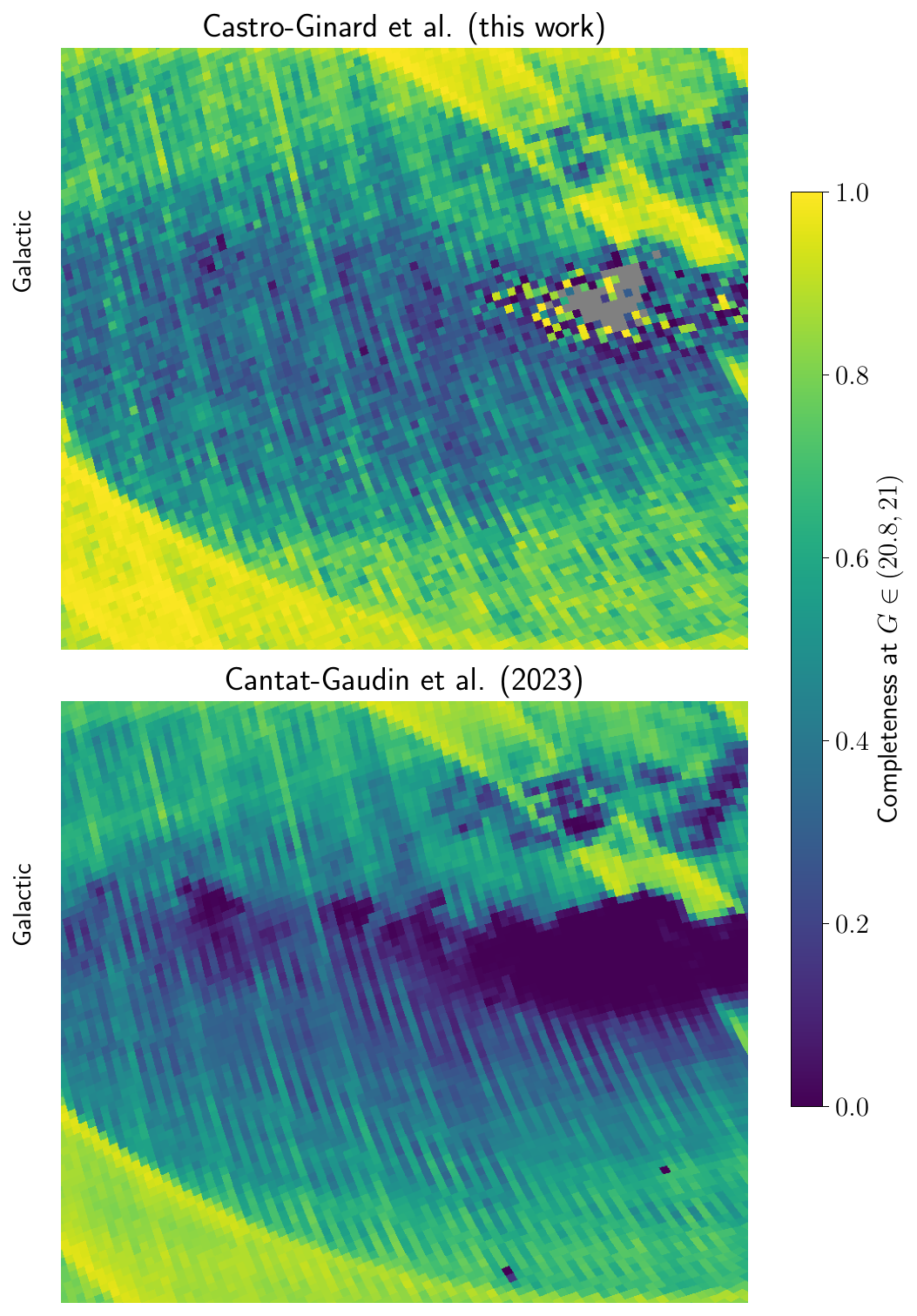}
    \caption{Side-by-side comparison of \gdr{3} selection function at $G\in(20.8,21)$ towards the Galactic centre, from this work (top) and from \citetalias[][bottom]{2023A&A...669A..55C}. Darker colours in both panels correspond to lower overall selection probability. Grey pixels in the top panel correspond to regions where \gaia does not report any source for the above magnitude interval.}
    \label{fig:galactic-centre}
\end{figure}

In Fig.~\ref{fig:galactic-centre}, we compare our model with the empirical selection function of \citetalias{2023A&A...669A..55C} in an overlapping region towards the Galactic centre for $G\in(20.8,21)$ to showcase the extreme faint end. Both methods trace the same overall scanning law and crowding features. The key difference lies in the treatment of missing data. While our model relies only on the \gaia catalogue, and therefore estimates the selection function only where \gaia data are available, \citetalias{2023A&A...669A..55C} assumes zero detections whenever \gaia has no coverage overlapping their ground truth survey (used for calibration). This pulls the \citetalias{2023A&A...669A..55C} selection probability toward lower values. The effect is most visible at faint magnitudes, where our model reports no estimate while \citetalias{2023A&A...669A..55C} report a selection probability close to zero, but it is present to some extent at every coordinate and magnitude.

\subsection{Completeness as a function of $G$}
\label{sec:completeness-G}

\begin{figure}
    \centering
    \includegraphics[width=\columnwidth]{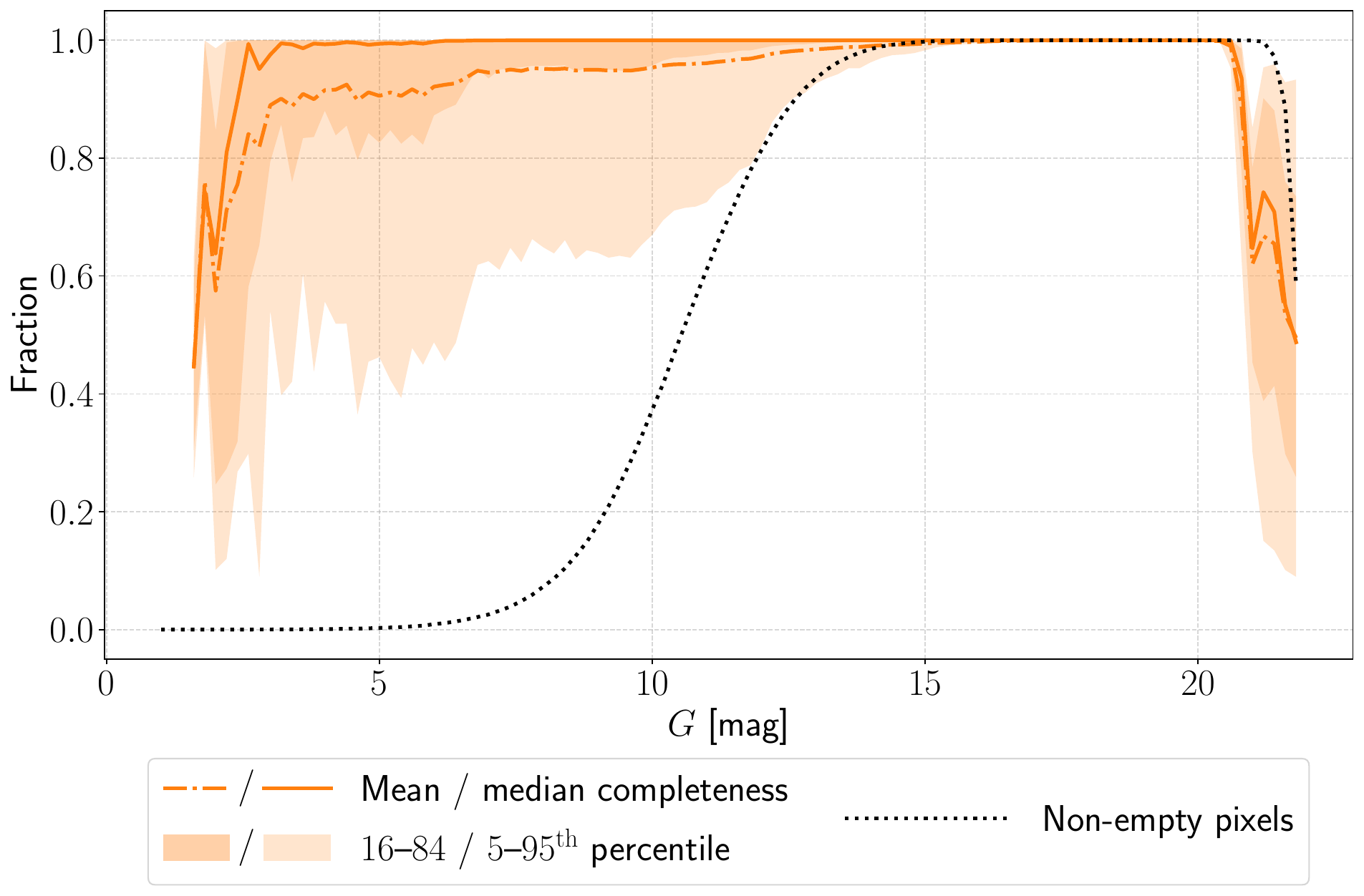}
    \caption{Sky-averaged selection function of the \gdr{3} catalogue as a function of $G$. The lines represent the on-sky mean (dash-dotted) and median (solid), with the shaded area showing the $5^{\rm th}$, $16^{\rm th}$, $84^{\rm th}$, and $95^{\rm th}$ percentiles. The black-dotted line shows the fraction of non-empty pixels contributing to the estimation of the selection function at each magnitude.}
    \label{fig:completeness-G}
\end{figure}

Our model is fully \gaia data-driven and estimates the selection function across the entire magnitude range for which \gaia data exists. This is an improvement with respect to \citetalias{2023A&A...669A..55C}, whose empirical model is only characterised near the faint end, and collapses by construction to a selection probability of $100\%$ or $0\%$ for brighter and fainter magnitudes, respectively. We show a sky-averaged selection probability in Fig.~\ref{fig:completeness-G}, where we confirm that the on-sky median shows \gdr{3} to be essentially complete across $G\in(5,19.5)$ except towards the Galactic centre, with the sky-averaged completeness remaining close to $1$ until $G\approx20.7$, after which it drops sharply. At magnitudes brighter than $G \sim 15$ the on-sky median selection probability remains close to $1$ but with larger scatter, since progressively fewer non-empty pixels contribute to the estimation.

\subsubsection*{Increased completeness at $G \gtrsim 21$}
\label{sec:increase-G21}

\begin{figure}
    \centering
    \includegraphics[width=0.8\columnwidth]{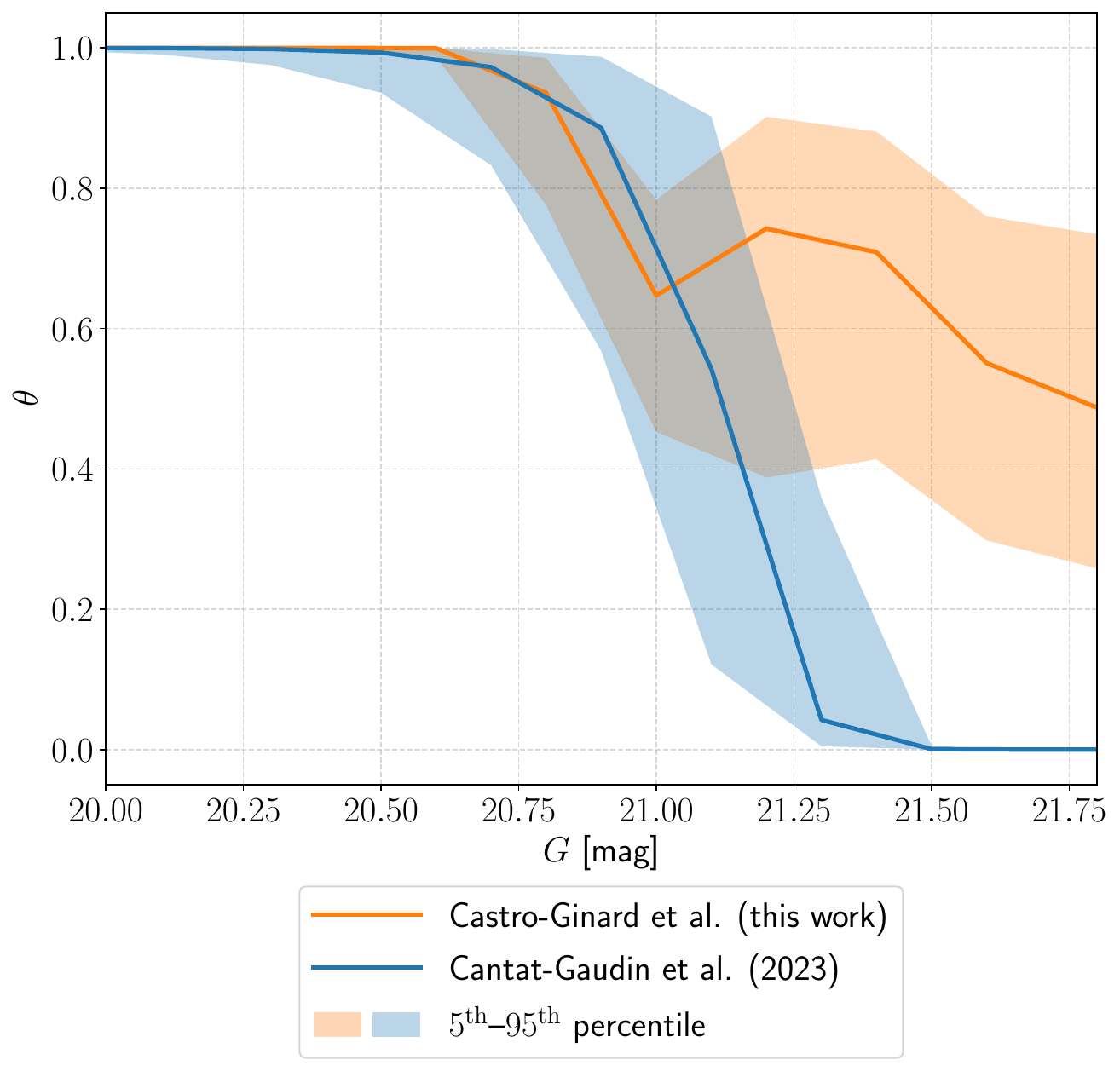}
    \caption{Comparison between our forward model (orange) and the empirical model of \citetalias[][blue]{2023A&A...669A..55C} at \gaia's faint end. In both cases, the solid line represents the on-sky median values and the shaded areas show the $5^{\rm th}$ and $95^{\rm th}$ percentiles.}
    \label{fig:comparison-G}
\end{figure}

We find a surprising increase in completeness at $G\gtrsim21$, beyond the nominal \gaia magnitude limit. Figure~\ref{fig:comparison-G} shows a zoomed-in version of Fig.~\ref{fig:completeness-G}, focusing on the very faint end and adding the \citetalias{2023A&A...669A..55C} selection function for comparison. Both estimations agree closely from $G = 20$ down to $G\approx21$, with differences being attributable to the different models and assumptions. Beyond this point, the two diverge, with the empirical selection probability of \citetalias{2023A&A...669A..55C} monotonically falling to $0$ by model construction, and our forward model rising to a local maximum at $G\approx21.2$ before slowly decreasing again. We attribute this difference to the simpler model in \citetalias{2023A&A...669A..55C}, where only the position of the completeness decline and how sharply it falls are free parameters, and the selection probability is forced to either $1$ or $0$ outside this regime. 

\begin{figure}
    \centering
    \includegraphics[width=\columnwidth]{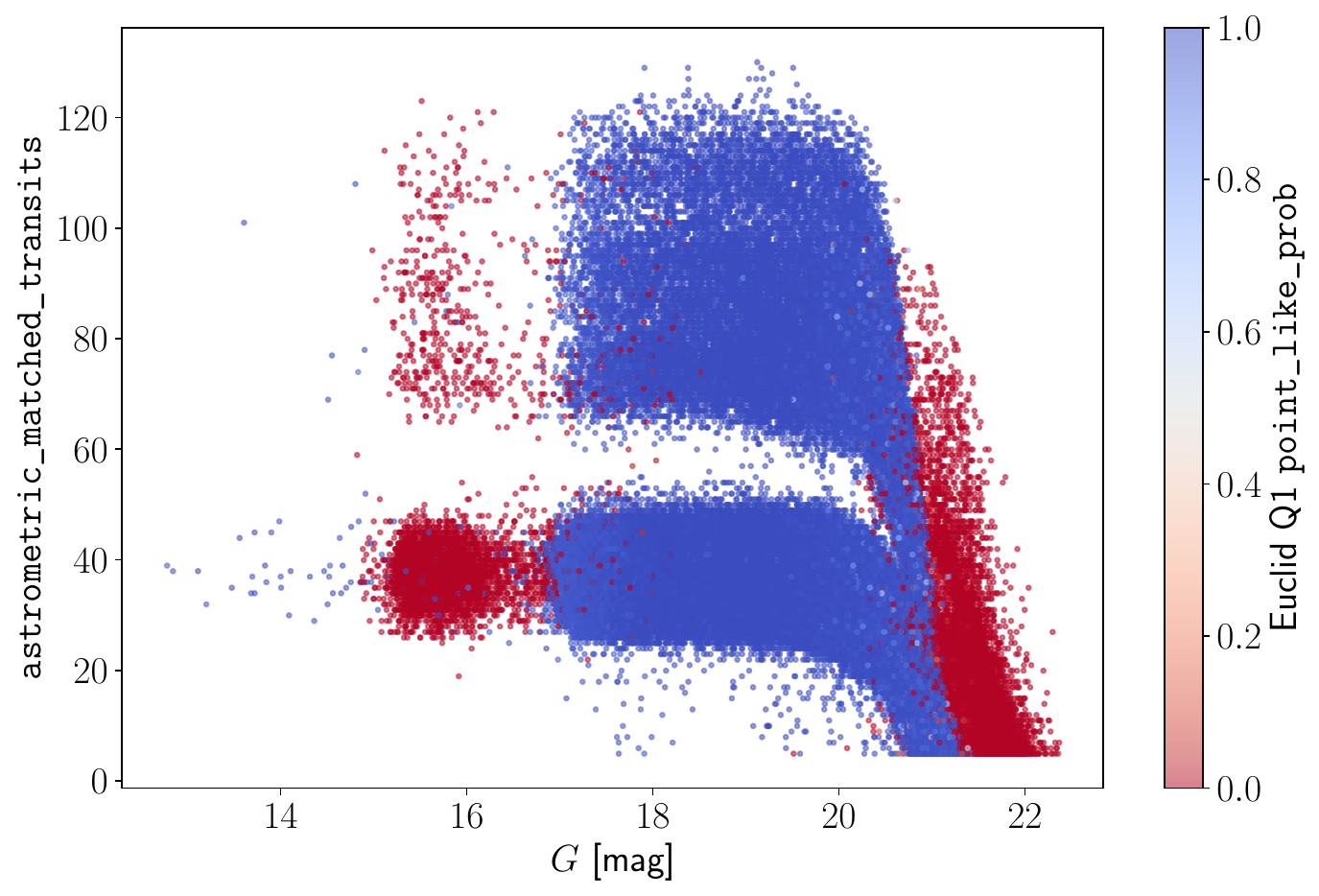}
    \caption{Number of \gdr{3} \fieldName{astrometric_matched_transits} as a function of $G$, for sources in common between \gdr{3} and the Euclid Q1 MER catalogue. The sources are colour-coded by their probability of being a point-like source. The population of extended sources (in red) corresponds to the completeness increase seen in Fig.~\ref{fig:completeness-G}.}
    \label{fig:euclid-XM}
\end{figure}

The completeness increase beyond $G\approx21$ was already seen for \gdr{2} in \citetalias{2020MNRAS.497.4246B}, who opted to remove all sources at $G>21.3$ on the evidence that a subset of the faint sources had miscalibrated $G$ magnitudes in \gdr{2}. We did not discard sources fainter than $G\sim21$ but did investigate more closely the faintest sources that cause the peak in the selection at $G=21.2$. We attribute the completeness increase beyond $G\approx21$ to galaxies, or QSOs with detectable host galaxies, whose $G$ magnitude measured by \gaia is too faint because the observation windows are too small to contain all the flux from these extended sources. On-board \gaia, where the observation windows at the detection stage are larger, these sources are bright enough to be picked up. However, in the on-ground processing their treatment as point sources in the photometric data processing leads to underestimation of the total flux, and thus the sources get assigned $G$ magnitudes that are too faint. This leads to a group of sources that has a number of astrometric matched transits that is too high for their estimated magnitude, which inflates the inferred selection probability. We test this by cross-matching \gdr{3} with the Euclid Q1 MER final catalogue \citep{2026A&A...711A...4E}. \figrefalt{fig:euclid-XM} shows the number of \fieldName{astrometric_matched_transits} as a function of $G$ for the sources in common, colour-coded by the probability of being a point-like source (reported in Euclid Q1 as \fieldName{point_like_prob}). The excess of sources with a high number of matched transits at $G\gtrsim21$ corresponds to the low point-source probability population. This is consistent with the increase seen in \figref{fig:completeness-G} being driven by these miscalibrated extended sources rather than a genuine increase in \gaia's detection efficiency. We therefore recommend not using our selection function at $G>21$, where it overestimates the completeness for point sources.

\subsection{Release in GaiaUnlimited}
\label{sec:gaiaunlimited}

The selection function presented in this work is released in the context of the GaiaUnlimited project\footnote{\url{https://gaia-unlimited.org/}}, alongside other tools already available to estimate \gaia selection functions: i) an empirical model of the \gdr{3} source catalogue selection function \citepalias{2023A&A...669A..55C}, ii) a Bayesian method to estimate the selection function of \gaia sub-samples \citep{2023A&A...677A..37C}, iii) a characterisation of the unresolved binary detectability based on \texttt{RUWE} \citep{2024A&A...688A...1C}, and iv) a joint \gaia-APOGEE selection function \citep{2024A&A...683A.128C}. These are released together as a single, pip-installable Python package\footnote{\url{https://github.com/gaia-unlimited/gaiaunlimited}}, with full online documentation\footnote{\url{https://gaiaunlimited.readthedocs.io/en/latest/index.html}} including tutorials that showcase their application to specific science cases, such as the old stellar disc traced by red clump stars \citep{2025A&A...701A.270K} and the density of metal-poor giants in the inner Galaxy \citep{2024arXiv241022250K}.

For the selection function derived in this work, GaiaUnlimited provides a dedicated class that returns the selection probability of \equref{eq:selprob} for any given celestial position and $G$ magnitude, together with the underlying effective scanning law of \secref{sec:effective-scanning-law}. We show in Appendix~\ref{app:gaiaunlimited-example} a worked example to query both the selection probability and the effective scanning law.

\section{Summary and Conclusions}
\label{sec:conclusions}

In this paper, we presented a hierarchical Bayesian model of the \gdr{3} source catalogue selection function, fit directly to every DR3 source with a reported celestial position, $G$ magnitude, and \fieldName{astrometric_matched_transits}. Following the truncated-Binomial approach of \citetalias{2020MNRAS.497.4246B}, we model the detection efficiency $\theta$ per HEALPix and magnitude bin as a Beta distribution, but depart from their previous implementation in how we treat the number of scans $n$. Rather than reconstructing $n$ from the catalogue, which pulls $n$ towards $k$, the number of detections, and therefore inflates $\theta$, the detection efficiency, we infer the effective number of scans $N_{\rm eff}$ as a latent variable per HEALPix, shared across all magnitude bins and informed by GOST and the known data-taking gaps. In this way, the reduced detection efficiency of faint sources in crowded fields is absorbed directly into the fitted $\theta$, requiring no separate, explicit crowding correction term.

This design choice results in an effective scanning law per HEALPix at a given resolution (level $7$ in this paper), instead of the per-source output provided by GOST. The recovered $N_{\rm eff}$ is statistically well constrained across most of the sky, with the posterior width at zero (one) for $\sim 90\%$ ($10\%$) of the HEALPix, degrading to a width of $8$ scans for the least constrained pixel. The resulting selection function is essentially complete in $G \in (15,19.5)$ everywhere except the Galactic centre, where crowding already limits \gaia selection probability to $50\%$ at $G \sim 19.5$, and its on-sky median remains close to $1$ down to $G\sim5$. From $G\sim19.5$, the incompleteness widens and deepens with magnitude, and by $G \sim 21$ the scanning law pattern becomes the dominant feature on the sky (see \figref{fig:magnitude-slices}).

Comparing our selection function against the two other all-sky estimates for \gdr{3}, we find good agreement on the general features with the empirical model of \citetalias{2023A&A...669A..55C}, but not with the forward model of \citetalias{2022MNRAS.509.6205E}. The key difference with respect to \citetalias{2023A&A...669A..55C} lies in how the two models handle missing data, with \citetalias{2023A&A...669A..55C} assuming zero detections wherever their calibration survey has no overlapping \gaia sources. This pulls their selection probability towards low values at the faint end, precisely where our model reports no estimate instead. The two selection functions closely agree across most of the overlapping magnitude range, but diverge exactly where the functional form of \citetalias{2023A&A...669A..55C} collapses to either $0$ or $1$ by construction, outside the regime where it was calibrated. Our model formally extends over the full magnitude range for which \gaia data exists, but at $G\lesssim15$ the small number of sources per pixel and magnitude bin leaves the selection function poorly constrained, and quantifying it at the bright end at our chosen resolution (HEALPix level $7$ and $0.2$\,mag bins) remains an open problem. Because $N_{\rm eff}$ is already constrained by the fainter sources, what is missing is information on the detection efficiency, which could be gained by pooling sources in larger HEALPix or wider magnitude bins at the bright end, or by hierarchically coupling $(A,B)$ across neighbouring pixels and magnitude bins. The forward model of \citetalias{2022MNRAS.509.6205E}, instead, is systematically more complete than both, reaching $100\%$ at $G\sim21$ across most of the sky. We attribute this to its detection efficiency, fitted to \gdr{2} with a scan count reconstructed from the catalogue itself, and then applied to \edr{3} scan counts that are also lower than our $N_{\rm eff}$.

Our model results in an artificial increase in completeness at $G\gtrsim21$ which is driven by extended sources (galaxies) for which the $G$ magnitude listed in \gdr{3} is fainter than their true, point-source-equivalent magnitude. This results in more matched transits for the fainter magnitude, inflating the inferred selection probability. A future improvement could be to filter out sources for which magnitudes are likely overestimated due to being extended. For now we would advise not to use the selection function beyond $G=21$. 

Our selection function does not depend explicitly on colour. Colour-dependent variations in the detection efficiency at fixed $G$ are absorbed into the width of the Beta distribution (\secref{sec:effmodel}), so $\mathcal{S}_{\textit{Gaia}}$ represents an average over the colours of the sources in each pixel and magnitude bin. Adding colour as a further dimension is possible but would restrict the fit to sources with $G_{\rm BP}$ and $G_{\rm RP}$ photometry.

\gdr{4} will roughly double the observational baseline from $34$ to $66$ months, giving every celestial position more scans to constrain the same hierarchical model. The \gdr{4} \fieldName{gaia_source} table will merge outputs of different processing pipelines. However, our model generalises directly to any \gaia source with a reported matched transit count, since it depends only on celestial position, magnitude and this count, regardless of which pipeline produced it.

\begin{acknowledgements}

This work is a result of the GaiaUnlimited project, which has received funding from the European Union’s Horizon 2020 research and innovation program under grant agreement No 101004110. The GaiaUnlimited project was started at the 2019 Santa Barbara Gaia Sprint, hosted by the Kavli Institute for Theoretical Physics at the University of California, Santa Barbara.

ACG acknowledges support from Horizon Europe HORIZON-CL4-2023-SPACE-01-71 project funded under grant agreement no. 101135205.

This work has made use of results from the European Space Agency (ESA) space mission {\it Gaia}, the data from which were processed by the {\it Gaia Data Processing and Analysis Consortium} (DPAC).  Funding for the DPAC has been provided by national institutions, in particular, the institutions participating in the {\it Gaia} Multilateral Agreement. The {\it Gaia} mission website is \url{http://www.cosmos.esa.int/gaia}. The authors are current or past members of the ESA {\it Gaia} mission team and of the {\it Gaia} DPAC.

This work has made use of the Euclid Q1 data from the Euclid mission of the European Space Agency (ESA), 2025, \url{https://doi.org/10.57780/esa-2853f3b}.

This work made use of the following software: 
Astropy, a community-developed core Python package for Astronomy \citep[\url{http://www.astropy.org}]{astropy:2013, astropy:2018, astropy:2022},
astroquery, an astronomical web-querying package in Python \citep[\url{https://astroquery.readthedocs.io/en/latest/}]{2019AJ....157...98G},
IPython \citep[\url{https://ipython.org/}]{ipython:2007},
Jupyter (\url{https://jupyter.org/}), 
Matplotlib \citep[\url{https://matplotlib.org}]{matplotlib:2007},
SciPy \citep[\url{https://www.scipy.org}]{scipy:2020}, 
NumPy \citep[\url{https://numpy.org}]{numpy:2020},
healpy \citep[\url{https://healpy.readthedocs.io}]{2019JOSS....4.1298Z},
and TOPCAT \citep[\url{http://www.starlink.ac.uk/topcat/}]{topcat:2005}.

This research has made use of the Science Explorer, funded by NASA under Cooperative Agreement 80NSSC21M00561.

\end{acknowledgements}

\bibliographystyle{aa}
\bibliography{bibliography}

@ARTICLE{2018A&A...616A...1G,
   author = {{Gaia Collaboration} and {Brown}, A.~G.~A. and {Vallenari}, A. and 
	{Prusti}, T. and {de Bruijne}, J.~H.~J. and {Babusiaux}, C. and 
	{Bailer-Jones}, C.~A.~L. and {Biermann}, M. and {Evans}, D.~W. and 
	{Eyer}, L. and et al.},
    title = "{Gaia Data Release 2. Summary of the contents and survey properties}",
  journal = {\aap},
archivePrefix = "arXiv",
   eprint = {1804.09365},
     year = 2018,
    month = aug,
   volume = 616,
      eid = {A1},
    pages = {A1},
      doi = {10.1051/0004-6361/201833051},
   adsurl = {http://adsabs.harvard.edu/abs/2018A%26A...616A...1G}
}

@ARTICLE{2016A&A...595A...1G,
   author = {{Gaia Collaboration} and {Prusti}, T. and {de Bruijne}, J.~H.~J. and 
	{Brown}, A.~G.~A. and {Vallenari}, A. and {Babusiaux}, C. and 
	{Bailer-Jones}, C.~A.~L. and {Bastian}, U. and {Biermann}, M. and 
	{Evans}, D.~W. and et al.},
    title = "{The Gaia mission}",
  journal = {\aap},
archivePrefix = "arXiv",
   eprint = {1609.04153},
 primaryClass = "astro-ph.IM",
     year = 2016,
    month = nov,
   volume = 595,
      eid = {A1},
    pages = {A1},
      doi = {10.1051/0004-6361/201629272},
   adsurl = {http://adsabs.harvard.edu/abs/2016A%26A...595A...1G}
}

@ARTICLE{2021A&A...649A...1G,
       author = {{Gaia Collaboration} and {Brown}, A.~G.~A. and {Vallenari}, A. and {Prusti}, T. and {de Bruijne}, J.~H.~J. and {Babusiaux}, C. and {Biermann}, M. and {Creevey}, O.~L. and {Evans}, D.~W. and {Eyer}, L. and {Hutton}, A. and {Jansen}, F. and {Jordi}, C. and {Klioner}, S.~A. and {Lammers}, U. and {Lindegren}, L. and {Luri}, X. and {Mignard}, F. and {Panem}, C. and {Pourbaix}, D. and {Randich}, S. and {Sartoretti}, P. and {Soubiran}, C. and {Walton}, N.~A. and {Arenou}, F. and {Bailer-Jones}, C.~A.~L. and {Bastian}, U. and {Cropper}, M. and {Drimmel}, R. and {Katz}, D. and {Lattanzi}, M.~G. and {van Leeuwen}, F. and {Bakker}, J. and {Cacciari}, C. and {Casta{\~n}eda}, J. and {De Angeli}, F. and {Ducourant}, C. and {Fabricius}, C. and {Fouesneau}, M. and {Fr{\'e}mat}, Y. and {Guerra}, R. and {Guerrier}, A. and {Guiraud}, J. and {Jean-Antoine Piccolo}, A. and {Masana}, E. and {Messineo}, R. and {Mowlavi}, N. and {Nicolas}, C. and {Nienartowicz}, K. and {Pailler}, F. and {Panuzzo}, P. and {Riclet}, F. and {Roux}, W. and {Seabroke}, G.~M. and {Sordo}, R. and {Tanga}, P. and {Th{\'e}venin}, F. and {Gracia-Abril}, G. and {Portell}, J. and {Teyssier}, D. and {Altmann}, M. and {Andrae}, R. and {Bellas-Velidis}, I. and {Benson}, K. and {Berthier}, J. and {Blomme}, R. and {Brugaletta}, E. and {Burgess}, P.~W. and {Busso}, G. and {Carry}, B. and {Cellino}, A. and {Cheek}, N. and {Clementini}, G. and {Damerdji}, Y. and {Davidson}, M. and {Delchambre}, L. and {Dell'Oro}, A. and {Fern{\'a}ndez-Hern{\'a}ndez}, J. and {Galluccio}, L. and {Garc{\'\i}a-Lario}, P. and {Garcia-Reinaldos}, M. and {Gonz{\'a}lez-N{\'u}{\~n}ez}, J. and {Gosset}, E. and {Haigron}, R. and {Halbwachs}, J. -L. and {Hambly}, N.~C. and {Harrison}, D.~L. and {Hatzidimitriou}, D. and {Heiter}, U. and {Hern{\'a}ndez}, J. and {Hestroffer}, D. and {Hodgkin}, S.~T. and {Holl}, B. and {Jan{\ss}en}, K. and {Jevardat de Fombelle}, G. and {Jordan}, S. and {Krone-Martins}, A. and {Lanzafame}, A.~C. and {L{\"o}ffler}, W. and {Lorca}, A. and {Manteiga}, M. and {Marchal}, O. and {Marrese}, P.~M. and {Moitinho}, A. and {Mora}, A. and {Muinonen}, K. and {Osborne}, P. and {Pancino}, E. and {Pauwels}, T. and {Petit}, J. -M. and {Recio-Blanco}, A. and {Richards}, P.~J. and {Riello}, M. and {Rimoldini}, L. and {Robin}, A.~C. and {Roegiers}, T. and {Rybizki}, J. and {Sarro}, L.~M. and {Siopis}, C. and {Smith}, M. and {Sozzetti}, A. and {Ulla}, A. and {Utrilla}, E. and {van Leeuwen}, M. and {van Reeven}, W. and {Abbas}, U. and {Abreu Aramburu}, A. and {Accart}, S. and {Aerts}, C. and {Aguado}, J.~J. and {Ajaj}, M. and {Altavilla}, G. and {{\'A}lvarez}, M.~A. and {{\'A}lvarez Cid-Fuentes}, J. and {Alves}, J. and {Anderson}, R.~I. and {Anglada Varela}, E. and {Antoja}, T. and {Audard}, M. and {Baines}, D. and {Baker}, S.~G. and {Balaguer-N{\'u}{\~n}ez}, L. and {Balbinot}, E. and {Balog}, Z. and {Barache}, C. and {Barbato}, D. and {Barros}, M. and {Barstow}, M.~A. and {Bartolom{\'e}}, S. and {Bassilana}, J. -L. and {Bauchet}, N. and {Baudesson-Stella}, A. and {Becciani}, U. and {Bellazzini}, M. and {Bernet}, M. and {Bertone}, S. and {Bianchi}, L. and {Blanco-Cuaresma}, S. and {Boch}, T. and {Bombrun}, A. and {Bossini}, D. and {Bouquillon}, S. and {Bragaglia}, A. and {Bramante}, L. and {Breedt}, E. and {Bressan}, A. and {Brouillet}, N. and {Bucciarelli}, B. and {Burlacu}, A. and {Busonero}, D. and {Butkevich}, A.~G. and {Buzzi}, R. and {Caffau}, E. and {Cancelliere}, R. and {C{\'a}novas}, H. and {Cantat-Gaudin}, T. and {Carballo}, R. and {Carlucci}, T. and {Carnerero}, M.~I. and {Carrasco}, J.~M. and {Casamiquela}, L. and {Castellani}, M. and {Castro-Ginard}, A. and {Castro Sampol}, P. and {Chaoul}, L. and {Charlot}, P. and {Chemin}, L. and {Chiavassa}, A. and {Cioni}, M. -R.~L. and {Comoretto}, G. and {Cooper}, W.~J. and {Cornez}, T. and {Cowell}, S. and {Crifo}, F. and {Crosta}, M. and {Crowley}, C. and {Dafonte}, C. and {Dapergolas}, A. and {David}, M. and {David}, P. and {de Laverny}, P. and {De Luise}, F. and {De March}, R. and {De Ridder}, J. and {de Souza}, R. and {de Teodoro}, P. and {de Torres}, A. and {del Peloso}, E.~F. and {del Pozo}, E. and {Delbo}, M. and {Delgado}, A. and {Delgado}, H.~E. and {Delisle}, J. -B. and {Di Matteo}, P. and {Diakite}, S. and {Diener}, C. and {Distefano}, E. and {Dolding}, C. and {Eappachen}, D. and {Edvardsson}, B. and {Enke}, H. and {Esquej}, P. and {Fabre}, C. and {Fabrizio}, M. and {Faigler}, S. and {Fedorets}, G. and {Fernique}, P. and {Fienga}, A. and {Figueras}, F. and {Fouron}, C. and {Fragkoudi}, F. and {Fraile}, E. and {Franke}, F. and {Gai}, M. and {Garabato}, D. and {Garcia-Gutierrez}, A. and {Garc{\'\i}a-Torres}, M. and {Garofalo}, A. and {Gavras}, P. and {Gerlach}, E. and {Geyer}, R. and {Giacobbe}, P. and {Gilmore}, G. and {Girona}, S. and {Giuffrida}, G. and {Gomel}, R. and {Gomez}, A. and {Gonzalez-Santamaria}, I. and {Gonz{\'a}lez-Vidal}, J.~J. and {Granvik}, M. and {Guti{\'e}rrez-S{\'a}nchez}, R. and {Guy}, L.~P. and {Hauser}, M. and {Haywood}, M. and {Helmi}, A. and {Hidalgo}, S.~L. and {Hilger}, T. and {H{\l}adczuk}, N. and {Hobbs}, D. and {Holland}, G. and {Huckle}, H.~E. and {Jasniewicz}, G. and {Jonker}, P.~G. and {Juaristi Campillo}, J. and {Julbe}, F. and {Karbevska}, L. and {Kervella}, P. and {Khanna}, S. and {Kochoska}, A. and {Kontizas}, M. and {Kordopatis}, G. and {Korn}, A.~J. and {Kostrzewa-Rutkowska}, Z. and {Kruszy{\'n}ska}, K. and {Lambert}, S. and {Lanza}, A.~F. and {Lasne}, Y. and {Le Campion}, J. -F. and {Le Fustec}, Y. and {Lebreton}, Y. and {Lebzelter}, T. and {Leccia}, S. and {Leclerc}, N. and {Lecoeur-Taibi}, I. and {Liao}, S. and {Licata}, E. and {Lindstr{\o}m}, E.~P. and {Lister}, T.~A. and {Livanou}, E. and {Lobel}, A. and {Madrero Pardo}, P. and {Managau}, S. and {Mann}, R.~G. and {Marchant}, J.~M. and {Marconi}, M. and {Marcos Santos}, M.~M.~S. and {Marinoni}, S. and {Marocco}, F. and {Marshall}, D.~J. and {Martin Polo}, L. and {Mart{\'\i}n-Fleitas}, J.~M. and {Masip}, A. and {Massari}, D. and {Mastrobuono-Battisti}, A. and {Mazeh}, T. and {McMillan}, P.~J. and {Messina}, S. and {Michalik}, D. and {Millar}, N.~R. and {Mints}, A. and {Molina}, D. and {Molinaro}, R. and {Moln{\'a}r}, L. and {Montegriffo}, P. and {Mor}, R. and {Morbidelli}, R. and {Morel}, T. and {Morris}, D. and {Mulone}, A.~F. and {Munoz}, D. and {Muraveva}, T. and {Murphy}, C.~P. and {Musella}, I. and {Noval}, L. and {Ord{\'e}novic}, C. and {Orr{\`u}}, G. and {Osinde}, J. and {Pagani}, C. and {Pagano}, I. and {Palaversa}, L. and {Palicio}, P.~A. and {Panahi}, A. and {Pawlak}, M. and {Pe{\~n}alosa Esteller}, X. and {Penttil{\"a}}, A. and {Piersimoni}, A.~M. and {Pineau}, F. -X. and {Plachy}, E. and {Plum}, G. and {Poggio}, E. and {Poretti}, E. and {Poujoulet}, E. and {Pr{\v{s}}a}, A. and {Pulone}, L. and {Racero}, E. and {Ragaini}, S. and {Rainer}, M. and {Raiteri}, C.~M. and {Rambaux}, N. and {Ramos}, P. and {Ramos-Lerate}, M. and {Re Fiorentin}, P. and {Regibo}, S. and {Reyl{\'e}}, C. and {Ripepi}, V. and {Riva}, A. and {Rixon}, G. and {Robichon}, N. and {Robin}, C. and {Roelens}, M. and {Rohrbasser}, L. and {Romero-G{\'o}mez}, M. and {Rowell}, N. and {Royer}, F. and {Rybicki}, K.~A. and {Sadowski}, G. and {Sagrist{\`a} Sell{\'e}s}, A. and {Sahlmann}, J. and {Salgado}, J. and {Salguero}, E. and {Samaras}, N. and {Sanchez Gimenez}, V. and {Sanna}, N. and {Santove{\~n}a}, R. and {Sarasso}, M. and {Schultheis}, M. and {Sciacca}, E. and {Segol}, M. and {Segovia}, J.~C. and {S{\'e}gransan}, D. and {Semeux}, D. and {Shahaf}, S. and {Siddiqui}, H.~I. and {Siebert}, A. and {Siltala}, L. and {Slezak}, E. and {Smart}, R.~L. and {Solano}, E. and {Solitro}, F. and {Souami}, D. and {Souchay}, J. and {Spagna}, A. and {Spoto}, F. and {Steele}, I.~A. and {Steidelm{\"u}ller}, H. and {Stephenson}, C.~A. and {S{\"u}veges}, M. and {Szabados}, L. and {Szegedi-Elek}, E. and {Taris}, F. and {Tauran}, G. and {Taylor}, M.~B. and {Teixeira}, R. and {Thuillot}, W. and {Tonello}, N. and {Torra}, F. and {Torra}, J. and {Turon}, C. and {Unger}, N. and {Vaillant}, M. and {van Dillen}, E. and {Vanel}, O. and {Vecchiato}, A. and {Viala}, Y. and {Vicente}, D. and {Voutsinas}, S. and {Weiler}, M. and {Wevers}, T. and {Wyrzykowski}, {\L}. and {Yoldas}, A. and {Yvard}, P. and {Zhao}, H. and {Zorec}, J. and {Zucker}, S. and {Zurbach}, C. and {Zwitter}, T.},
        title = "{Gaia Early Data Release 3. Summary of the contents and survey properties}",
      journal = {\aap},
         year = 2021,
        month = may,
       volume = {649},
          eid = {A1},
        pages = {A1},
          doi = {10.1051/0004-6361/202039657},
archivePrefix = {arXiv},
       eprint = {2012.01533},
 primaryClass = {astro-ph.GA},
       adsurl = {https://ui.adsabs.harvard.edu/abs/2021A&A...649A...1G}
}

@ARTICLE{2021A&A...649A...2L,
       author = {{Lindegren}, L. and {Klioner}, S.~A. and {Hern{\'a}ndez}, J. and {Bombrun}, A. and {Ramos-Lerate}, M. and {Steidelm{\"u}ller}, H. and {Bastian}, U. and {Biermann}, M. and {de Torres}, A. and {Gerlach}, E. and {Geyer}, R. and {Hilger}, T. and {Hobbs}, D. and {Lammers}, U. and {McMillan}, P.~J. and {Stephenson}, C.~A. and {Casta{\~n}eda}, J. and {Davidson}, M. and {Fabricius}, C. and {Gracia-Abril}, G. and {Portell}, J. and {Rowell}, N. and {Teyssier}, D. and {Torra}, F. and {Bartolom{\'e}}, S. and {Clotet}, M. and {Garralda}, N. and {Gonz{\'a}lez-Vidal}, J.~J. and {Torra}, J. and {Abbas}, U. and {Altmann}, M. and {Anglada Varela}, E. and {Balaguer-N{\'u}{\~n}ez}, L. and {Balog}, Z. and {Barache}, C. and {Becciani}, U. and {Bernet}, M. and {Bertone}, S. and {Bianchi}, L. and {Bouquillon}, S. and {Brown}, A.~G.~A. and {Bucciarelli}, B. and {Busonero}, D. and {Butkevich}, A.~G. and {Buzzi}, R. and {Cancelliere}, R. and {Carlucci}, T. and {Charlot}, P. and {Cioni}, M. -R.~L. and {Crosta}, M. and {Crowley}, C. and {del Peloso}, E.~F. and {del Pozo}, E. and {Drimmel}, R. and {Esquej}, P. and {Fienga}, A. and {Fraile}, E. and {Gai}, M. and {Garcia-Reinaldos}, M. and {Guerra}, R. and {Hambly}, N.~C. and {Hauser}, M. and {Jan{\ss}en}, K. and {Jordan}, S. and {Kostrzewa-Rutkowska}, Z. and {Lattanzi}, M.~G. and {Liao}, S. and {Licata}, E. and {Lister}, T.~A. and {L{\"o}ffler}, W. and {Marchant}, J.~M. and {Masip}, A. and {Mignard}, F. and {Mints}, A. and {Molina}, D. and {Mora}, A. and {Morbidelli}, R. and {Murphy}, C.~P. and {Pagani}, C. and {Panuzzo}, P. and {Pe{\~n}alosa Esteller}, X. and {Poggio}, E. and {Re Fiorentin}, P. and {Riva}, A. and {Sagrist{\`a} Sell{\'e}s}, A. and {Sanchez Gimenez}, V. and {Sarasso}, M. and {Sciacca}, E. and {Siddiqui}, H.~I. and {Smart}, R.~L. and {Souami}, D. and {Spagna}, A. and {Steele}, I.~A. and {Taris}, F. and {Utrilla}, E. and {van Reeven}, W. and {Vecchiato}, A.},
        title = "{Gaia Early Data Release 3. The astrometric solution}",
      journal = {\aap},
         year = 2021,
        month = may,
       volume = {649},
          eid = {A2},
        pages = {A2},
          doi = {10.1051/0004-6361/202039709},
archivePrefix = {arXiv},
       eprint = {2012.03380},
 primaryClass = {astro-ph.IM},
       adsurl = {https://ui.adsabs.harvard.edu/abs/2021A&A...649A...2L}
}

@ARTICLE{2020MNRAS.497.4246B,
       author = {{Boubert}, Douglas and {Everall}, Andrew},
        title = "{Completeness of the Gaia verse II: what are the odds that a star is missing from Gaia DR2?}",
      journal = {\mnras},
         year = 2020,
        month = oct,
       volume = {497},
       number = {4},
        pages = {4246-4261},
          doi = {10.1093/mnras/staa2305},
archivePrefix = {arXiv},
       eprint = {2005.08983},
 primaryClass = {astro-ph.GA},
       adsurl = {https://ui.adsabs.harvard.edu/abs/2020MNRAS.497.4246B}
}

@ARTICLE{2023A&A...669A..55C,
       author = {{Cantat-Gaudin}, Tristan and {Fouesneau}, Morgan and {Rix}, Hans-Walter and {Brown}, Anthony G.~A. and {Castro-Ginard}, Alfred and {Kostrzewa-Rutkowska}, Zuzanna and {Drimmel}, Ronald and {Hogg}, David W. and {Casey}, Andrew R. and {Khanna}, Shourya and {Oh}, Semyeong and {Price-Whelan}, Adrian M. and {Belokurov}, Vasily and {Saydjari}, Andrew K. and {Green}, G.},
        title = "{An empirical model of the Gaia DR3 selection function}",
      journal = {\aap},
         year = 2023,
        month = jan,
       volume = {669},
          eid = {A55},
        pages = {A55},
          doi = {10.1051/0004-6361/202244784},
archivePrefix = {arXiv},
       eprint = {2208.09335},
 primaryClass = {astro-ph.GA},
       adsurl = {https://ui.adsabs.harvard.edu/abs/2023A&A...669A..55C}
}

@ARTICLE{2022MNRAS.509.6205E,
       author = {{Everall}, Andrew and {Boubert}, Douglas},
        title = "{Completeness of the Gaia verse - V. Astrometry and radial velocity sample selection functions in Gaia EDR3}",
      journal = {\mnras},
         year = 2022,
        month = feb,
       volume = {509},
       number = {4},
        pages = {6205-6224},
          doi = {10.1093/mnras/stab3262},
archivePrefix = {arXiv},
       eprint = {2111.04127},
 primaryClass = {astro-ph.GA},
       adsurl = {https://ui.adsabs.harvard.edu/abs/2022MNRAS.509.6205E}
}

@ARTICLE{2021AJ....162..142R,
       author = {{Rix}, Hans-Walter and {Hogg}, David W. and {Boubert}, Douglas and {Brown}, Anthony G.~A. and {Casey}, Andrew and {Drimmel}, Ronald and {Everall}, Andrew and {Fouesneau}, Morgan and {Price-Whelan}, Adrian M.},
        title = "{Selection Functions in Astronomical Data Modeling, with the Space Density of White Dwarfs as a Worked Example}",
      journal = {\aj},
         year = 2021,
        month = oct,
       volume = {162},
       number = {4},
          eid = {142},
        pages = {142},
          doi = {10.3847/1538-3881/ac0c13},
archivePrefix = {arXiv},
       eprint = {2106.07653},
 primaryClass = {astro-ph.IM},
       adsurl = {https://ui.adsabs.harvard.edu/abs/2021AJ....162..142R}
}

@ARTICLE{2023A&A...674A...1G,
       author = {{Gaia Collaboration} and {Vallenari}, A. and {Brown}, A.~G.~A. and {Prusti}, T. and {de Bruijne}, J.~H.~J. and {Arenou}, F. and {Babusiaux}, C. and {Biermann}, M. and {Creevey}, O.~L. and {Ducourant}, C. and {Evans}, D.~W. and {Eyer}, L. and {Guerra}, R. and {Hutton}, A. and {Jordi}, C. and {Klioner}, S.~A. and {Lammers}, U.~L. and {Lindegren}, L. and {Luri}, X. and {Mignard}, F. and {Panem}, C. and {Pourbaix}, D. and {Randich}, S. and {Sartoretti}, P. and {Soubiran}, C. and {Tanga}, P. and {Walton}, N.~A. and {Bailer-Jones}, C.~A.~L. and {Bastian}, U. and {Drimmel}, R. and {Jansen}, F. and {Katz}, D. and {Lattanzi}, M.~G. and {van Leeuwen}, F. and {Bakker}, J. and {Cacciari}, C. and {Casta{\~n}eda}, J. and {De Angeli}, F. and {Fabricius}, C. and {Fouesneau}, M. and {Fr{\'e}mat}, Y. and {Galluccio}, L. and {Guerrier}, A. and {Heiter}, U. and {Masana}, E. and {Messineo}, R. and {Mowlavi}, N. and {Nicolas}, C. and {Nienartowicz}, K. and {Pailler}, F. and {Panuzzo}, P. and {Riclet}, F. and {Roux}, W. and {Seabroke}, G.~M. and {Sordo}, R. and {Th{\'e}venin}, F. and {Gracia-Abril}, G. and {Portell}, J. and {Teyssier}, D. and {Altmann}, M. and {Andrae}, R. and {Audard}, M. and {Bellas-Velidis}, I. and {Benson}, K. and {Berthier}, J. and {Blomme}, R. and {Burgess}, P.~W. and {Busonero}, D. and {Busso}, G. and {C{\'a}novas}, H. and {Carry}, B. and {Cellino}, A. and {Cheek}, N. and {Clementini}, G. and {Damerdji}, Y. and {Davidson}, M. and {de Teodoro}, P. and {Nu{\~n}ez Campos}, M. and {Delchambre}, L. and {Dell'Oro}, A. and {Esquej}, P. and {Fern{\'a}ndez-Hern{\'a}ndez}, J. and {Fraile}, E. and {Garabato}, D. and {Garc{\'\i}a-Lario}, P. and {Gosset}, E. and {Haigron}, R. and {Halbwachs}, J. -L. and {Hambly}, N.~C. and {Harrison}, D.~L. and {Hern{\'a}ndez}, J. and {Hestroffer}, D. and {Hodgkin}, S.~T. and {Holl}, B. and {Jan{\ss}en}, K. and {Jevardat de Fombelle}, G. and {Jordan}, S. and {Krone-Martins}, A. and {Lanzafame}, A.~C. and {L{\"o}ffler}, W. and {Marchal}, O. and {Marrese}, P.~M. and {Moitinho}, A. and {Muinonen}, K. and {Osborne}, P. and {Pancino}, E. and {Pauwels}, T. and {Recio-Blanco}, A. and {Reyl{\'e}}, C. and {Riello}, M. and {Rimoldini}, L. and {Roegiers}, T. and {Rybizki}, J. and {Sarro}, L.~M. and {Siopis}, C. and {Smith}, M. and {Sozzetti}, A. and {Utrilla}, E. and {van Leeuwen}, M. and {Abbas}, U. and {{\'A}brah{\'a}m}, P. and {Abreu Aramburu}, A. and {Aerts}, C. and {Aguado}, J.~J. and {Ajaj}, M. and {Aldea-Montero}, F. and {Altavilla}, G. and {{\'A}lvarez}, M.~A. and {Alves}, J. and {Anders}, F. and {Anderson}, R.~I. and {Anglada Varela}, E. and {Antoja}, T. and {Baines}, D. and {Baker}, S.~G. and {Balaguer-N{\'u}{\~n}ez}, L. and {Balbinot}, E. and {Balog}, Z. and {Barache}, C. and {Barbato}, D. and {Barros}, M. and {Barstow}, M.~A. and {Bartolom{\'e}}, S. and {Bassilana}, J. -L. and {Bauchet}, N. and {Becciani}, U. and {Bellazzini}, M. and {Berihuete}, A. and {Bernet}, M. and {Bertone}, S. and {Bianchi}, L. and {Binnenfeld}, A. and {Blanco-Cuaresma}, S. and {Blazere}, A. and {Boch}, T. and {Bombrun}, A. and {Bossini}, D. and {Bouquillon}, S. and {Bragaglia}, A. and {Bramante}, L. and {Breedt}, E. and {Bressan}, A. and {Brouillet}, N. and {Brugaletta}, E. and {Bucciarelli}, B. and {Burlacu}, A. and {Butkevich}, A.~G. and {Buzzi}, R. and {Caffau}, E. and {Cancelliere}, R. and {Cantat-Gaudin}, T. and {Carballo}, R. and {Carlucci}, T. and {Carnerero}, M.~I. and {Carrasco}, J.~M. and {Casamiquela}, L. and {Castellani}, M. and {Castro-Ginard}, A. and {Chaoul}, L. and {Charlot}, P. and {Chemin}, L. and {Chiaramida}, V. and {Chiavassa}, A. and {Chornay}, N. and {Comoretto}, G. and {Contursi}, G. and {Cooper}, W.~J. and {Cornez}, T. and {Cowell}, S. and {Crifo}, F. and {Cropper}, M. and {Crosta}, M. and {Crowley}, C. and {Dafonte}, C. and {Dapergolas}, A. and {David}, M. and {David}, P. and {de Laverny}, P. and {De Luise}, F. and {De March}, R. and {De Ridder}, J. and {de Souza}, R. and {de Torres}, A. and {del Peloso}, E.~F. and {del Pozo}, E. and {Delbo}, M. and {Delgado}, A. and {Delisle}, J. -B. and {Demouchy}, C. and {Dharmawardena}, T.~E. and {Di Matteo}, P. and {Diakite}, S. and {Diener}, C. and {Distefano}, E. and {Dolding}, C. and {Edvardsson}, B. and {Enke}, H. and {Fabre}, C. and {Fabrizio}, M. and {Faigler}, S. and {Fedorets}, G. and {Fernique}, P. and {Fienga}, A. and {Figueras}, F. and {Fournier}, Y. and {Fouron}, C. and {Fragkoudi}, F. and {Gai}, M. and {Garcia-Gutierrez}, A. and {Garcia-Reinaldos}, M. and {Garc{\'\i}a-Torres}, M. and {Garofalo}, A. and {Gavel}, A. and {Gavras}, P. and {Gerlach}, E. and {Geyer}, R. and {Giacobbe}, P. and {Gilmore}, G. and {Girona}, S. and {Giuffrida}, G. and {Gomel}, R. and {Gomez}, A. and {Gonz{\'a}lez-N{\'u}{\~n}ez}, J. and {Gonz{\'a}lez-Santamar{\'\i}a}, I. and {Gonz{\'a}lez-Vidal}, J.~J. and {Granvik}, M. and {Guillout}, P. and {Guiraud}, J. and {Guti{\'e}rrez-S{\'a}nchez}, R. and {Guy}, L.~P. and {Hatzidimitriou}, D. and {Hauser}, M. and {Haywood}, M. and {Helmer}, A. and {Helmi}, A. and {Sarmiento}, M.~H. and {Hidalgo}, S.~L. and {Hilger}, T. and {H{\l}adczuk}, N. and {Hobbs}, D. and {Holland}, G. and {Huckle}, H.~E. and {Jardine}, K. and {Jasniewicz}, G. and {Jean-Antoine Piccolo}, A. and {Jim{\'e}nez-Arranz}, {\'O}. and {Jorissen}, A. and {Juaristi Campillo}, J. and {Julbe}, F. and {Karbevska}, L. and {Kervella}, P. and {Khanna}, S. and {Kontizas}, M. and {Kordopatis}, G. and {Korn}, A.~J. and {K{\'o}sp{\'a}l}, {\'A}. and {Kostrzewa-Rutkowska}, Z. and {Kruszy{\'n}ska}, K. and {Kun}, M. and {Laizeau}, P. and {Lambert}, S. and {Lanza}, A.~F. and {Lasne}, Y. and {Le Campion}, J. -F. and {Lebreton}, Y. and {Lebzelter}, T. and {Leccia}, S. and {Leclerc}, N. and {Lecoeur-Taibi}, I. and {Liao}, S. and {Licata}, E.~L. and {Lindstr{\o}m}, H.~E.~P. and {Lister}, T.~A. and {Livanou}, E. and {Lobel}, A. and {Lorca}, A. and {Loup}, C. and {Madrero Pardo}, P. and {Magdaleno Romeo}, A. and {Managau}, S. and {Mann}, R.~G. and {Manteiga}, M. and {Marchant}, J.~M. and {Marconi}, M. and {Marcos}, J. and {Marcos Santos}, M.~M.~S. and {Mar{\'\i}n Pina}, D. and {Marinoni}, S. and {Marocco}, F. and {Marshall}, D.~J. and {Martin Polo}, L. and {Mart{\'\i}n-Fleitas}, J.~M. and {Marton}, G. and {Mary}, N. and {Masip}, A. and {Massari}, D. and {Mastrobuono-Battisti}, A. and {Mazeh}, T. and {McMillan}, P.~J. and {Messina}, S. and {Michalik}, D. and {Millar}, N.~R. and {Mints}, A. and {Molina}, D. and {Molinaro}, R. and {Moln{\'a}r}, L. and {Monari}, G. and {Mongui{\'o}}, M. and {Montegriffo}, P. and {Montero}, A. and {Mor}, R. and {Mora}, A. and {Morbidelli}, R. and {Morel}, T. and {Morris}, D. and {Muraveva}, T. and {Murphy}, C.~P. and {Musella}, I. and {Nagy}, Z. and {Noval}, L. and {Oca{\~n}a}, F. and {Ogden}, A. and {Ordenovic}, C. and {Osinde}, J.~O. and {Pagani}, C. and {Pagano}, I. and {Palaversa}, L. and {Palicio}, P.~A. and {Pallas-Quintela}, L. and {Panahi}, A. and {Payne-Wardenaar}, S. and {Pe{\~n}alosa Esteller}, X. and {Penttil{\"a}}, A. and {Pichon}, B. and {Piersimoni}, A.~M. and {Pineau}, F. -X. and {Plachy}, E. and {Plum}, G. and {Poggio}, E. and {Pr{\v{s}}a}, A. and {Pulone}, L. and {Racero}, E. and {Ragaini}, S. and {Rainer}, M. and {Raiteri}, C.~M. and {Rambaux}, N. and {Ramos}, P. and {Ramos-Lerate}, M. and {Re Fiorentin}, P. and {Regibo}, S. and {Richards}, P.~J. and {Rios Diaz}, C. and {Ripepi}, V. and {Riva}, A. and {Rix}, H. -W. and {Rixon}, G. and {Robichon}, N. and {Robin}, A.~C. and {Robin}, C. and {Roelens}, M. and {Rogues}, H.~R.~O. and {Rohrbasser}, L. and {Romero-G{\'o}mez}, M. and {Rowell}, N. and {Royer}, F. and {Ruz Mieres}, D. and {Rybicki}, K.~A. and {Sadowski}, G. and {S{\'a}ez N{\'u}{\~n}ez}, A. and {Sagrist{\`a} Sell{\'e}s}, A. and {Sahlmann}, J. and {Salguero}, E. and {Samaras}, N. and {Sanchez Gimenez}, V. and {Sanna}, N. and {Santove{\~n}a}, R. and {Sarasso}, M. and {Schultheis}, M. and {Sciacca}, E. and {Segol}, M. and {Segovia}, J.~C. and {S{\'e}gransan}, D. and {Semeux}, D. and {Shahaf}, S. and {Siddiqui}, H.~I. and {Siebert}, A. and {Siltala}, L. and {Silvelo}, A. and {Slezak}, E. and {Slezak}, I. and {Smart}, R.~L. and {Snaith}, O.~N. and {Solano}, E. and {Solitro}, F. and {Souami}, D. and {Souchay}, J. and {Spagna}, A. and {Spina}, L. and {Spoto}, F. and {Steele}, I.~A. and {Steidelm{\"u}ller}, H. and {Stephenson}, C.~A. and {S{\"u}veges}, M. and {Surdej}, J. and {Szabados}, L. and {Szegedi-Elek}, E. and {Taris}, F. and {Taylor}, M.~B. and {Teixeira}, R. and {Tolomei}, L. and {Tonello}, N. and {Torra}, F. and {Torra}, J. and {Torralba Elipe}, G. and {Trabucchi}, M. and {Tsounis}, A.~T. and {Turon}, C. and {Ulla}, A. and {Unger}, N. and {Vaillant}, M.~V. and {van Dillen}, E. and {van Reeven}, W. and {Vanel}, O. and {Vecchiato}, A. and {Viala}, Y. and {Vicente}, D. and {Voutsinas}, S. and {Weiler}, M. and {Wevers}, T. and {Wyrzykowski}, {\L}. and {Yoldas}, A. and {Yvard}, P. and {Zhao}, H. and {Zorec}, J. and {Zucker}, S. and {Zwitter}, T.},
        title = "{Gaia Data Release 3. Summary of the content and survey properties}",
      journal = {\aap},
         year = 2023,
        month = jun,
       volume = {674},
          eid = {A1},
        pages = {A1},
          doi = {10.1051/0004-6361/202243940},
archivePrefix = {arXiv},
       eprint = {2208.00211},
 primaryClass = {astro-ph.GA},
       adsurl = {https://ui.adsabs.harvard.edu/abs/2023A&A...674A...1G}
}

@ARTICLE{2023A&A...677A..37C,
       author = {{Castro-Ginard}, Alfred and {Brown}, Anthony G.~A. and {Kostrzewa-Rutkowska}, Zuzanna and {Cantat-Gaudin}, Tristan and {Drimmel}, Ronald and {Oh}, Semyeong and {Belokurov}, Vasily and {Casey}, Andrew R. and {Fouesneau}, Morgan and {Khanna}, Shourya and {Price-Whelan}, Adrian M. and {Rix}, Hans-Walter},
        title = "{Estimating the selection function of Gaia DR3 subsamples}",
      journal = {\aap},
         year = 2023,
        month = sep,
       volume = {677},
          eid = {A37},
        pages = {A37},
          doi = {10.1051/0004-6361/202346547},
archivePrefix = {arXiv},
       eprint = {2303.17738},
 primaryClass = {astro-ph.GA},
       adsurl = {https://ui.adsabs.harvard.edu/abs/2023A&A...677A..37C}
}

@ARTICLE{2024A&A...683A.128C,
       author = {{Cantat-Gaudin}, Tristan and {Fouesneau}, Morgan and {Rix}, Hans-Walter and {Brown}, Anthony G.~A. and {Drimmel}, Ronald and {Castro-Ginard}, Alfred and {Khanna}, Shourya and {Belokurov}, Vasily and {Casey}, Andrew R.},
        title = "{Uniting Gaia and APOGEE to unveil the cosmic chemistry of the Milky Way disc}",
      journal = {\aap},
         year = 2024,
        month = mar,
       volume = {683},
          eid = {A128},
        pages = {A128},
          doi = {10.1051/0004-6361/202348018},
archivePrefix = {arXiv},
       eprint = {2401.05023},
 primaryClass = {astro-ph.GA},
       adsurl = {https://ui.adsabs.harvard.edu/abs/2024A&A...683A.128C}
}

@ARTICLE{2005ApJ...622..759G,
       author = {{G{\'o}rski}, K.~M. and {Hivon}, E. and {Banday}, A.~J. and {Wandelt}, B.~D. and {Hansen}, F.~K. and {Reinecke}, M. and {Bartelmann}, M.},
        title = "{HEALPix: A Framework for High-Resolution Discretization and Fast Analysis of Data Distributed on the Sphere}",
      journal = {\apj},
         year = 2005,
        month = apr,
       volume = {622},
       number = {2},
        pages = {759-771},
          doi = {10.1086/427976},
archivePrefix = {arXiv},
       eprint = {astro-ph/0409513},
 primaryClass = {astro-ph},
       adsurl = {https://ui.adsabs.harvard.edu/abs/2005ApJ...622..759G}
}

@ARTICLE{2013PASP..125..306F,
       author = {{Foreman-Mackey}, Daniel and {Hogg}, David W. and {Lang}, Dustin and {Goodman}, Jonathan},
        title = "{emcee: The MCMC Hammer}",
      journal = {\pasp},
         year = 2013,
        month = mar,
       volume = {125},
       number = {925},
        pages = {306},
          doi = {10.1086/670067},
archivePrefix = {arXiv},
       eprint = {1202.3665},
 primaryClass = {astro-ph.IM},
       adsurl = {https://ui.adsabs.harvard.edu/abs/2013PASP..125..306F}
}

@ARTICLE{2026A&A...711A...4E,
       author = {{Euclid Collaboration} and {Romelli}, E. and {K{\"u}mmel}, M. and {Dole}, H. and {Gracia-Carpio}, J. and {Merlin}, E. and {Galeotta}, S. and {Fang}, Y. and {Castellano}, M. and {Caro}, F. and {Soubrie}, E. and {Maurin}, L. and {Cabanac}, R. and {Dimauro}, P. and {Huertas-Company}, M. and {Lepinzan}, M.~D. and {Vassallo}, T. and {Walmsley}, M. and {Zinchenko}, I.~A. and {Boucaud}, A. and {Calabro}, A. and {Roscani}, V. and {Tramacere}, A. and {Douspis}, M. and {Fontana}, A. and {Aghanim}, N. and {Altieri}, B. and {Amara}, A. and {Andreon}, S. and {Auricchio}, N. and {Aussel}, H. and {Baccigalupi}, C. and {Baldi}, M. and {Balestra}, A. and {Bardelli}, S. and {Basset}, A. and {Battaglia}, P. and {Belikov}, A.~N. and {Biviano}, A. and {Bonchi}, A. and {Bonino}, D. and {Branchini}, E. and {Brescia}, M. and {Brinchmann}, J. and {Camera}, S. and {Ca{\~n}as-Herrera}, G. and {Capobianco}, V. and {Carbone}, C. and {Carretero}, J. and {Casas}, S. and {Castander}, F.~J. and {Castignani}, G. and {Cavuoti}, S. and {Chambers}, K.~C. and {Cimatti}, A. and {Colodro-Conde}, C. and {Congedo}, G. and {Conselice}, C.~J. and {Conversi}, L. and {Copin}, Y. and {Courbin}, F. and {Courtois}, H.~M. and {Cropper}, M. and {Cuby}, J.-G. and {Da Silva}, A. and {da Silva}, R. and {Degaudenzi}, H. and {de la Torre}, S. and {De Lucia}, G. and {Di Giorgio}, A.~M. and {Dubath}, F. and {Duncan}, C.~A.~J. and {Dupac}, X. and {Dusini}, S. and {Escoffier}, S. and {Fabricius}, M. and {Farina}, M. and {Farinelli}, R. and {Faustini}, F. and {Ferriol}, S. and {Finelli}, F. and {Fotopoulou}, S. and {Fourmanoit}, N. and {Frailis}, M. and {Franceschi}, E. and {George}, K. and {Gillard}, W. and {Gillis}, B. and {Giocoli}, C. and {Granett}, B.~R. and {Grazian}, A. and {Grupp}, F. and {Gwyn}, S. and {Haugan}, S.~V.~H. and {Hoar}, J. and {Hoekstra}, H. and {Holmes}, W. and {Hook}, I.~M. and {Hormuth}, F. and {Hornstrup}, A. and {Jahnke}, K. and {Jhabvala}, M. and {Joachimi}, B. and {Keih{\"a}nen}, E. and {Kermiche}, S. and {Kiessling}, A. and {Kilbinger}, M. and {Kubik}, B. and {Kuijken}, K. and {Kunz}, M. and {Kurki-Suonio}, H. and {Laureijs}, R. and {Le Boulc'h}, Q. and {Le Brun}, A.~M.~C. and {Le Mignant}, D. and {Liebing}, P. and {Ligori}, S. and {Lilje}, P.~B. and {Lindholm}, V. and {Lloro}, I. and {Mainetti}, G. and {Maino}, D. and {Maiorano}, E. and {Mansutti}, O. and {Marcin}, S. and {Marggraf}, O. and {Martinelli}, M. and {Martinet}, N. and {Marulli}, F. and {Massey}, R. and {Maurogordato}, S. and {McCracken}, H.~J. and {Medinaceli}, E. and {Mei}, S. and {Melchior}, M. and {Mellier}, Y. and {Meneghetti}, M. and {Meylan}, G. and {Mora}, A. and {Moresco}, M. and {Moscardini}, L. and {Nakajima}, R. and {Neissner}, C. and {Nichol}, R.~C. and {Niemi}, S.-M. and {Nightingale}, J.~W. and {Padilla}, C. and {Paltani}, S. and {Pasian}, F. and {Pedersen}, K. and {Percival}, W.~J. and {Pettorino}, V. and {Pires}, S. and {Polenta}, G. and {Poncet}, M. and {Popa}, L.~A. and {Pozzetti}, L. and {Racca}, G.~D. and {Raison}, F. and {Renzi}, A. and {Rhodes}, J. and {Riccio}, G. and {Roncarelli}, M. and {Rossetti}, E. and {Saglia}, R. and {Sakr}, Z. and {S{\'a}nchez}, A.~G. and {Sapone}, D. and {Sartoris}, B. and {Schewtschenko}, J.~A. and {Schirmer}, M. and {Schneider}, P. and {Scodeggio}, M. and {Secroun}, A. and {Sefusatti}, E. and {Seidel}, G. and {Seiffert}, M. and {Serrano}, S. and {Simon}, P. and {Sirignano}, C. and {Sirri}, G. and {Spurio Mancini}, A. and {Stanco}, L. and {Steinwagner}, J. and {Tallada-Cresp{\'\i}}, P. and {Tavagnacco}, D. and {Taylor}, A.~N. and {Teplitz}, H.~I. and {Tereno}, I. and {Tessore}, N. and {Toft}, S. and {Toledo-Moreo}, R. and {Torradeflot}, F. and {Tsyganov}, A. and {Tutusaus}, I. and {Valentijn}, E.~A. and {Valenziano}, L. and {Valiviita}, J. and {Verdoes Kleijn}, G. and {Veropalumbo}, A.},
        title = "{Euclid Quick Data Release (Q1): IV. From images to multi-wavelength catalogues: The Euclid MERge processing function}",
      journal = {\aap},
         year = 2026,
        month = jun,
       volume = {711},
          eid = {A4},
        pages = {A4},
          doi = {10.1051/0004-6361/202554586},
archivePrefix = {arXiv},
       eprint = {2503.15305},
 primaryClass = {astro-ph.IM},
       adsurl = {https://ui.adsabs.harvard.edu/abs/2026A&A...711A...4E}
}

@ARTICLE{2024A&A...688A...1C,
       author = {{Castro-Ginard}, Alfred and {Penoyre}, Zephyr and {Casey}, Andrew R. and {Brown}, Anthony G.~A. and {Belokurov}, Vasily and {Cantat-Gaudin}, Tristan and {Drimmel}, Ronald and {Fouesneau}, Morgan and {Khanna}, Shourya and {Kurbatov}, Evgeny P. and {Price-Whelan}, Adrian M. and {Rix}, Hans-Walter and {Smart}, Richard L.},
        title = "{Gaia DR3 detectability of unresolved binary systems}",
      journal = {\aap},
         year = 2024,
        month = aug,
       volume = {688},
          eid = {A1},
        pages = {A1},
          doi = {10.1051/0004-6361/202450172},
archivePrefix = {arXiv},
       eprint = {2404.14127},
 primaryClass = {astro-ph.GA},
       adsurl = {https://ui.adsabs.harvard.edu/abs/2024A&A...688A...1C}
}

@ARTICLE{2025A&A...701A.270K,
       author = {{Khanna}, Shourya and {Yu}, Jie and {Drimmel}, Ronald and {Poggio}, Eloisa and {Cantat-Gaudin}, Tristan and {Castro-Ginard}, Alfred and {Kurbatov}, Evgeny and {Belokurov}, Vasily and {Brown}, Anthony and {Fouesneau}, Morgan and {Casey}, Andrew and {Rix}, Hans-Walter},
        title = "{GaiaUnlimited: The old stellar disc of the Milky Way as traced by the red clump}",
      journal = {\aap},
         year = 2025,
        month = sep,
       volume = {701},
          eid = {A270},
        pages = {A270},
          doi = {10.1051/0004-6361/202452798},
archivePrefix = {arXiv},
       eprint = {2410.22036},
 primaryClass = {astro-ph.GA},
       adsurl = {https://ui.adsabs.harvard.edu/abs/2025A&A...701A.270K}
}

@ARTICLE{2024arXiv241022250K,
       author = {{Kurbatov}, Evgeny P. and {Belokurov}, Vasily and {Koposov}, Sergey and {Kravtsov}, Andrey and {Davies}, Elliot Y. and {Brown}, Anthony G.~A. and {Cantat-Gaudin}, Tristan and {Castro-Ginard}, Alfred and {Casey}, Andrew R. and {Drimmel}, Ronald and {Fouesneau}, Morgan and {Khanna}, Shourya and {Rix}, Hans-Walter and {Wallace}, Alex},
        title = "{The realm of Aurora. Density distribution of metal-poor giants in the heart of the Galaxy}",
      journal = {arXiv e-prints},
         year = 2024,
        month = oct,
          eid = {arXiv:2410.22250},
        pages = {arXiv:2410.22250},
          doi = {10.48550/arXiv.2410.22250},
archivePrefix = {arXiv},
       eprint = {2410.22250},
 primaryClass = {astro-ph.GA},
       adsurl = {https://ui.adsabs.harvard.edu/abs/2024arXiv241022250K}
}

@ARTICLE{1922MeLuF.100....1M,
       author = {{Malmquist}, K.~G.},
        title = "{On some relations in stellar statistics}",
      journal = {Meddelanden fran Lunds Astronomiska Observatorium Serie I},
         year = 1922,
        month = mar,
       volume = {100},
        pages = {1-52},
       adsurl = {https://ui.adsabs.harvard.edu/abs/1922MeLuF.100....1M}
}

@article{astropy:2013,
        Adsurl = {https://adsabs.harvard.edu/abs/2013A%26A...558A..33A},
        Archiveprefix = {arXiv},
        Author = {{Astropy Collaboration} and {Robitaille}, T.~P. and {Tollerud}, E.~J. and {Greenfield}, P. and {Droettboom}, M. and {Bray}, E. and {Aldcroft}, T. and {Davis}, M. and {Ginsburg}, A. and {Price-Whelan}, A.~M. and {Kerzendorf}, W.~E. and {Conley}, A. and {Crighton}, N. and {Barbary}, K. and {Muna}, D. and {Ferguson}, H. and {Grollier}, F. and {Parikh}, M.~M. and {Nair}, P.~H. and {Unther}, H.~M. and {Deil}, C. and {Woillez}, J. and {Conseil}, S. and {Kramer}, R. and {Turner}, J.~E.~H. and {Singer}, L. and {Fox}, R. and {Weaver}, B.~A. and {Zabalza}, V. and {Edwards}, Z.~I. and {Azalee Bostroem}, K. and {Burke}, D.~J. and {Casey}, A.~R. and {Crawford}, S.~M. and {Dencheva}, N. and {Ely}, J. and {Jenness}, T. and {Labrie}, K. and {Lim}, P.~L. and {Pierfederici}, F. and {Pontzen}, A. and {Ptak}, A. and {Refsdal}, B. and {Servillat}, M. and {Streicher}, O.},
        Doi = {10.1051/0004-6361/201322068},
        Eid = {A33},
        Eprint = {1307.6212},
        Journal = {\aap},
        Month = oct,
        Pages = {A33},
        Primaryclass = {astro-ph.IM},
        Title = {{Astropy: A community Python package for astronomy}},
        Volume = 558,
        Year = 2013}

@ARTICLE{astropy:2018,
               author = {{Astropy Collaboration} and {Price-Whelan}, A.~M. and
                 {Sip{\H{o}}cz}, B.~M. and {G{\"u}nther}, H.~M. and {Lim}, P.~L. and
                 {Crawford}, S.~M. and {Conseil}, S. and {Shupe}, D.~L. and
                 {Craig}, M.~W. and {Dencheva}, N. and {Ginsburg}, A. and {Vand
                erPlas}, J.~T. and {Bradley}, L.~D. and {P{\'e}rez-Su{\'a}rez}, D. and
                 {de Val-Borro}, M. and {Aldcroft}, T.~L. and {Cruz}, K.~L. and
                 {Robitaille}, T.~P. and {Tollerud}, E.~J. and {Ardelean}, C. and
                 {Babej}, T. and {Bach}, Y.~P. and {Bachetti}, M. and {Bakanov}, A.~V. and
                 {Bamford}, S.~P. and {Barentsen}, G. and {Barmby}, P. and
                 {Baumbach}, A. and {Berry}, K.~L. and {Biscani}, F. and {Boquien}, M. and
                 {Bostroem}, K.~A. and {Bouma}, L.~G. and {Brammer}, G.~B. and
                 {Bray}, E.~M. and {Breytenbach}, H. and {Buddelmeijer}, H. and
                 {Burke}, D.~J. and {Calderone}, G. and {Cano Rodr{\'\i}guez}, J.~L. and
                 {Cara}, M. and {Cardoso}, J.~V.~M. and {Cheedella}, S. and {Copin}, Y. and
                 {Corrales}, L. and {Crichton}, D. and {D'Avella}, D. and {Deil}, C. and
                 {Depagne}, {\'E}. and {Dietrich}, J.~P. and {Donath}, A. and
                 {Droettboom}, M. and {Earl}, N. and {Erben}, T. and {Fabbro}, S. and
                 {Ferreira}, L.~A. and {Finethy}, T. and {Fox}, R.~T. and
                 {Garrison}, L.~H. and {Gibbons}, S.~L.~J. and {Goldstein}, D.~A. and
                 {Gommers}, R. and {Greco}, J.~P. and {Greenfield}, P. and
                 {Groener}, A.~M. and {Grollier}, F. and {Hagen}, A. and {Hirst}, P. and
                 {Homeier}, D. and {Horton}, A.~J. and {Hosseinzadeh}, G. and {Hu}, L. and
                 {Hunkeler}, J.~S. and {Ivezi{\'c}}, {\v{Z}}. and {Jain}, A. and
                 {Jenness}, T. and {Kanarek}, G. and {Kendrew}, S. and {Kern}, N.~S. and
                 {Kerzendorf}, W.~E. and {Khvalko}, A. and {King}, J. and {Kirkby}, D. and
                 {Kulkarni}, A.~M. and {Kumar}, A. and {Lee}, A. and {Lenz}, D. and
                 {Littlefair}, S.~P. and {Ma}, Z. and {Macleod}, D.~M. and
                 {Mastropietro}, M. and {McCully}, C. and {Montagnac}, S. and
                 {Morris}, B.~M. and {Mueller}, M. and {Mumford}, S.~J. and {Muna}, D. and
                 {Murphy}, N.~A. and {Nelson}, S. and {Nguyen}, G.~H. and
                 {Ninan}, J.~P. and {N{\"o}the}, M. and {Ogaz}, S. and {Oh}, S. and
                 {Parejko}, J.~K. and {Parley}, N. and {Pascual}, S. and {Patil}, R. and
                 {Patil}, A.~A. and {Plunkett}, A.~L. and {Prochaska}, J.~X. and
                 {Rastogi}, T. and {Reddy Janga}, V. and {Sabater}, J. and
                 {Sakurikar}, P. and {Seifert}, M. and {Sherbert}, L.~E. and
                 {Sherwood-Taylor}, H. and {Shih}, A.~Y. and {Sick}, J. and
                 {Silbiger}, M.~T. and {Singanamalla}, S. and {Singer}, L.~P. and
                 {Sladen}, P.~H. and {Sooley}, K.~A. and {Sornarajah}, S. and
                 {Streicher}, O. and {Teuben}, P. and {Thomas}, S.~W. and
                 {Tremblay}, G.~R. and {Turner}, J.~E.~H. and {Terr{\'o}n}, V. and
                 {van Kerkwijk}, M.~H. and {de la Vega}, A. and {Watkins}, L.~L. and
                 {Weaver}, B.~A. and {Whitmore}, J.~B. and {Woillez}, J. and
                 {Zabalza}, V. and {Astropy Contributors}},
                title = "{The Astropy Project: Building an Open-science Project and Status of the v2.0 Core Package}",
              journal = {\aj},
                 year = 2018,
                month = sep,
               volume = {156},
               number = {3},
                  eid = {123},
                pages = {123},
                  doi = {10.3847/1538-3881/aabc4f},
        archivePrefix = {arXiv},
               eprint = {1801.02634},
         primaryClass = {astro-ph.IM},
               adsurl = {https://ui.adsabs.harvard.edu/abs/2018AJ....156..123A}
        }

@ARTICLE{astropy:2022,
               author = {{Astropy Collaboration} and {Price-Whelan}, Adrian M. and {Lim}, Pey Lian and {Earl}, Nicholas and {Starkman}, Nathaniel and {Bradley}, Larry and {Shupe}, David L. and {Patil}, Aarya A. and {Corrales}, Lia and {Brasseur}, C.~E. and {N{"o}the}, Maximilian and {Donath}, Axel and {Tollerud}, Erik and {Morris}, Brett M. and {Ginsburg}, Adam and {Vaher}, Eero and {Weaver}, Benjamin A. and {Tocknell}, James and {Jamieson}, William and {van Kerkwijk}, Marten H. and {Robitaille}, Thomas P. and {Merry}, Bruce and {Bachetti}, Matteo and {G{"u}nther}, H. Moritz and {Aldcroft}, Thomas L. and {Alvarado-Montes}, Jaime A. and {Archibald}, Anne M. and {B{'o}di}, Attila and {Bapat}, Shreyas and {Barentsen}, Geert and {Baz{'a}n}, Juanjo and {Biswas}, Manish and {Boquien}, M{'e}d{'e}ric and {Burke}, D.~J. and {Cara}, Daria and {Cara}, Mihai and {Conroy}, Kyle E. and {Conseil}, Simon and {Craig}, Matthew W. and {Cross}, Robert M. and {Cruz}, Kelle L. and {D'Eugenio}, Francesco and {Dencheva}, Nadia and {Devillepoix}, Hadrien A.~R. and {Dietrich}, J{"o}rg P. and {Eigenbrot}, Arthur Davis and {Erben}, Thomas and {Ferreira}, Leonardo and {Foreman-Mackey}, Daniel and {Fox}, Ryan and {Freij}, Nabil and {Garg}, Suyog and {Geda}, Robel and {Glattly}, Lauren and {Gondhalekar}, Yash and {Gordon}, Karl D. and {Grant}, David and {Greenfield}, Perry and {Groener}, Austen M. and {Guest}, Steve and {Gurovich}, Sebastian and {Handberg}, Rasmus and {Hart}, Akeem and {Hatfield-Dodds}, Zac and {Homeier}, Derek and {Hosseinzadeh}, Griffin and {Jenness}, Tim and {Jones}, Craig K. and {Joseph}, Prajwel and {Kalmbach}, J. Bryce and {Karamehmetoglu}, Emir and {Ka{l}uszy{'n}ski}, Miko{l}aj and {Kelley}, Michael S.~P. and {Kern}, Nicholas and {Kerzendorf}, Wolfgang E. and {Koch}, Eric W. and {Kulumani}, Shankar and {Lee}, Antony and {Ly}, Chun and {Ma}, Zhiyuan and {MacBride}, Conor and {Maljaars}, Jakob M. and {Muna}, Demitri and {Murphy}, N.~A. and {Norman}, Henrik and {O'Steen}, Richard and {Oman}, Kyle A. and {Pacifici}, Camilla and {Pascual}, Sergio and {Pascual-Granado}, J. and {Patil}, Rohit R. and {Perren}, Gabriel I. and {Pickering}, Timothy E. and {Rastogi}, Tanuj and {Roulston}, Benjamin R. and {Ryan}, Daniel F. and {Rykoff}, Eli S. and {Sabater}, Jose and {Sakurikar}, Parikshit and {Salgado}, Jes{'u}s and {Sanghi}, Aniket and {Saunders}, Nicholas and {Savchenko}, Volodymyr and {Schwardt}, Ludwig and {Seifert-Eckert}, Michael and {Shih}, Albert Y. and {Jain}, Anany Shrey and {Shukla}, Gyanendra and {Sick}, Jonathan and {Simpson}, Chris and {Singanamalla}, Sudheesh and {Singer}, Leo P. and {Singhal}, Jaladh and {Sinha}, Manodeep and {Sip{H{o}}cz}, Brigitta M. and {Spitler}, Lee R. and {Stansby}, David and {Streicher}, Ole and {{{S}}umak}, Jani and {Swinbank}, John D. and {Taranu}, Dan S. and {Tewary}, Nikita and {Tremblay}, Grant R. and {Val-Borro}, Miguel de and {Van Kooten}, Samuel J. and {Vasovi{'c}}, Zlatan and {Verma}, Shresth and {de Miranda Cardoso}, Jos{'e} Vin{'i}cius and {Williams}, Peter K.~G. and {Wilson}, Tom J. and {Winkel}, Benjamin and {Wood-Vasey}, W.~M. and {Xue}, Rui and {Yoachim}, Peter and {Zhang}, Chen and {Zonca}, Andrea and {Astropy Project Contributors}},
                title = "{The Astropy Project: Sustaining and Growing a Community-oriented Open-source Project and the Latest Major Release (v5.0) of the Core Package}",
              journal = {\apj},
                 year = 2022,
                month = aug,
               volume = {935},
               number = {2},
                  eid = {167},
                pages = {167},
                  doi = {10.3847/1538-4357/ac7c74},
        archivePrefix = {arXiv},
               eprint = {2206.14220},
         primaryClass = {astro-ph.IM},
               adsurl = {https://ui.adsabs.harvard.edu/abs/2022ApJ...935..167A}
        }

@ARTICLE{2019AJ....157...98G,
       author = {{Ginsburg}, Adam and {Sip{\H{o}}cz}, Brigitta M. and {Brasseur}, C.~E. and {Cowperthwaite}, Philip S. and {Craig}, Matthew W. and {Deil}, Christoph and {Guillochon}, James and {Guzman}, Giannina and {Liedtke}, Simon and {Lian Lim}, Pey and et al.},
        title = "{astroquery: An Astronomical Web-querying Package in Python}",
      journal = {\aj},
         year = 2019,
        month = mar,
       volume = {157},
       number = {3},
          eid = {98},
        pages = {98},
          doi = {10.3847/1538-3881/aafc33},
archivePrefix = {arXiv},
       eprint = {1901.04520},
 primaryClass = {astro-ph.IM},
       adsurl = {https://ui.adsabs.harvard.edu/abs/2019AJ....157...98G}
}

@Article{ipython:2007,
  Author    = {P\'erez, Fernando and Granger, Brian E.},
  Title     = {{IP}ython: a System for Interactive Scientific Computing},
  Journal   = {Computing in Science and Engineering},
  Volume    = {9},
  Number    = {3},
  Pages     = {21--29},
  month     = may,
  year      = 2007,
  url       = "http://ipython.org",
  ISSN      = "1521-9615",
  doi       = {10.1109/MCSE.2007.53},
  publisher = {IEEE Computer Society},
}

@INPROCEEDINGS{topcat:2005,
       author = {{Taylor}, M.~B.},
        title = "{TOPCAT \& STIL: Starlink Table/VOTable Processing Software}",
    booktitle = {Astronomical Data Analysis Software and Systems XIV},
         year = 2005,
       editor = {{Shopbell}, P. and {Britton}, M. and {Ebert}, R.},
       series = {Astronomical Society of the Pacific Conference Series},
       volume = {347},
        month = dec,
        pages = {29},
       adsurl = {https://ui.adsabs.harvard.edu/abs/2005ASPC..347...29T}
}

@ARTICLE{scipy:2020,
       author = {{Virtanen}, Pauli and {Gommers}, Ralf and {Oliphant},
         Travis E. and {Haberland}, Matt and {Reddy}, Tyler and
         {Cournapeau}, David and {Burovski}, Evgeni and {Peterson}, Pearu
         and {Weckesser}, Warren and {Bright}, Jonathan and {van der Walt},
         St{\'e}fan J.  and {Brett}, Matthew and {Wilson}, Joshua and
         {Jarrod Millman}, K.  and {Mayorov}, Nikolay and {Nelson}, Andrew
         R.~J. and {Jones}, Eric and {Kern}, Robert and {Larson}, Eric and
         {Carey}, CJ and {Polat}, {\.I}lhan and {Feng}, Yu and {Moore},
         Eric W. and {Vand erPlas}, Jake and {Laxalde}, Denis and
         {Perktold}, Josef and {Cimrman}, Robert and {Henriksen}, Ian and
         {Quintero}, E.~A. and {Harris}, Charles R and {Archibald}, Anne M.
         and {Ribeiro}, Ant{\^o}nio H. and {Pedregosa}, Fabian and
         {van Mulbregt}, Paul and {Contributors}, SciPy 1. 0},
        title = "{SciPy 1.0: Fundamental Algorithms for Scientific
                  Computing in Python}",
      journal = {Nature Methods},
      year = "2020",
      volume={17},
      pages={261--272},
      adsurl = {https://rdcu.be/b08Wh},
      doi = {https://doi.org/10.1038/s41592-019-0686-2},
}

@ARTICLE{numpy:2020,
  author  = {Harris, Charles R. and Millman, K. Jarrod and
            van der Walt, Stéfan J and Gommers, Ralf and
            Virtanen, Pauli and Cournapeau, David and
            Wieser, Eric and Taylor, Julian and Berg, Sebastian and
            Smith, Nathaniel J. and Kern, Robert and Picus, Matti and
            Hoyer, Stephan and van Kerkwijk, Marten H. and
            Brett, Matthew and Haldane, Allan and
            Fernández del Río, Jaime and Wiebe, Mark and
            Peterson, Pearu and Gérard-Marchant, Pierre and
            Sheppard, Kevin and Reddy, Tyler and Weckesser, Warren and
            Abbasi, Hameer and Gohlke, Christoph and
            Oliphant, Travis E.},
  title   = {Array programming with {NumPy}},
  journal = {Nature},
  year    = {2020},
  volume  = {585},
  pages   = {357–362},
  doi     = {10.1038/s41586-020-2649-2}
}

@ARTICLE{matplotlib:2007,
       author = {{Hunter}, John D.},
        title = "{Matplotlib: A 2D Graphics Environment}",
      journal = {Computing in Science and Engineering},
         year = 2007,
        month = may,
       volume = {9},
       number = {3},
        pages = {90-95},
          doi = {10.1109/MCSE.2007.55},
       adsurl = {https://ui.adsabs.harvard.edu/abs/2007CSE.....9...90H}
}

@ARTICLE{2015A&A...576A..74D,
       author = {{de Bruijne}, J.~H.~J. and {Allen}, M. and {Azaz}, S. and {Krone-Martins}, A. and {Prod'homme}, T. and {Hestroffer}, D.},
        title = "{Detecting stars, galaxies, and asteroids with Gaia}",
      journal = {\aap},
         year = 2015,
        month = apr,
       volume = {576},
          eid = {A74},
        pages = {A74},
          doi = {10.1051/0004-6361/201424018},
archivePrefix = {arXiv},
       eprint = {1502.00454},
 primaryClass = {astro-ph.IM},
       adsurl = {https://ui.adsabs.harvard.edu/abs/2015A&A...576A..74D}
}

@ARTICLE{2019JOSS....4.1298Z,
       author = {{Zonca}, Andrea and {Singer}, Leo and {Lenz}, Daniel and {Reinecke}, Martin and {Rosset}, Cyrille and {Hivon}, Eric and {Gorski}, Krzysztof},
        title = "{healpy: equal area pixelization and spherical harmonics transforms for data on the sphere in Python}",
      journal = {The Journal of Open Source Software},
         year = 2019,
        month = mar,
       volume = {4},
       number = {35},
          eid = {1298},
        pages = {1298},
          doi = {10.21105/joss.01298},
       adsurl = {https://ui.adsabs.harvard.edu/abs/2019JOSS....4.1298Z}
}

@ARTICLE{2018ApJS..234...39S,
       author = {{Schlafly}, E.~F. and {Green}, G.~M. and {Lang}, D. and {Daylan}, T. and {Finkbeiner}, D.~P. and {Lee}, A. and {Meisner}, A.~M. and {Schlegel}, D. and {Valdes}, F.},
        title = "{The DECam Plane Survey: Optical Photometry of Two Billion Objects in the Southern Galactic Plane}",
      journal = {\apjs},
         year = 2018,
        month = feb,
       volume = {234},
       number = {2},
          eid = {39},
        pages = {39},
          doi = {10.3847/1538-4365/aaa3e2},
archivePrefix = {arXiv},
       eprint = {1710.01309},
 primaryClass = {astro-ph.GA},
       adsurl = {https://ui.adsabs.harvard.edu/abs/2018ApJS..234...39S}
}

@INPROCEEDINGS{2022AAS...24031806S,
       author = {{Saydjari}, Andrew and {Schlafly}, Edward and {Lang}, Dustin and {Meisner}, Aaron and {Green}, Gregory and {Zucker}, Catherine and {Zelko}, Ioana and {Daylan}, Tansu and {Lee}, Albert and {Valdes}, Francisco and et al.},
        title = "{The DECam Plane Survey (DECaPS2): Optical photometry of 3.3 billion stars in the southern Galactic plane}",
    booktitle = {American Astronomical Society Meeting \#240},
         year = 2022,
       series = {American Astronomical Society Meeting Abstracts},
       volume = {240},
        month = jun,
          eid = {318.06},
        pages = {318.06},
       adsurl = {https://ui.adsabs.harvard.edu/abs/2022AAS...24031806S}
}

@InProceedings{10.1007/978-3-031-43943-8_1,
author="Jette, Morris A.
and Wickberg, Tim",
editor="Klus{\'a}{\v{c}}ek, Dalibor
and Corbal{\'a}n, Julita
and Rodrigo, Gonzalo P.",
title="Architecture of the Slurm Workload Manager",
booktitle="Job Scheduling Strategies for Parallel Processing",
year="2023",
publisher="Springer Nature Switzerland",
address="Cham",
pages="3--23",
isbn="978-3-031-43943-8"
}

\onecolumn
\begin{appendix}
\section{Using the GaiaUnlimited python package}
\label{app:gaiaunlimited-example}

The examples below provide the code to reproduce Fig.~\ref{fig:scanning-law} and Fig.~\ref{fig:magnitude-slices}. The data retrieved through these examples are also the source for Fig.~\ref{fig:galactic-centre}, Fig.~\ref{fig:completeness-G}, and Fig.~\ref{fig:comparison-G}. A more detailed tutorial on how to use the model presented in this paper is available on the GaiaUnlimited online documentation\footnote{\url{https://gaiaunlimited.readthedocs.io/en/latest/}}.

\begin{lstlisting}[language=Python, caption = {Code to retrieve and plot the effective scanning law at HEALPix level 7. The result is shown in Fig.~\ref{fig:scanning-law}.}, label = {lst:gaiaunlimited-scanning-law}]
from gaiaunlimited.selectionfunctions import DR3SelectionFunctionForwardModel
from gaiaunlimited.utils import get_healpix_centers
import healpy as hp
import matplotlib.pyplot as plt

# Initiate the selection function class
dr3_sf = DR3SelectionFunctionForwardModel()

# Query at coordinates for HEALPix level 7
coords_of_centers = get_healpix_centers(7) 

# Query the effective scanning law
scanning_law = dr3_sf.query_effective_scanning_law(coords_of_centers)

# Plot using the healpy package
hp.mollview(
        scanning_law['N_eff'],
        coord=["Celestial", "Galactic"],
        title=f"Scanning law",
    )
\end{lstlisting}

\begin{lstlisting}[language=Python, caption = {Code to retrieve and plot the all sky selection function at different magnitude bins. The result is shown in Fig.~\ref{fig:magnitude-slices}.}, label = {lst:magnitude-slices}]
from gaiaunlimited.selectionfunctions import DR3SelectionFunctionForwardModel
from gaiaunlimited.utils import get_healpix_centers
import numpy as np
import healpy as hp
import matplotlib.pyplot as plt

# Initiate the selection function class
dr3_sf = DR3SelectionFunctionForwardModel()

# Query at coordinates for HEALPix level 7 and magnitudes
coords_of_centers = get_healpix_centers(7) 
Gmags = np.array([19.5, 20.1, 20.7, 21])

# Query and plot the selection function
for g in Gmags:
    prob_map = dr3_sf.query(coords_of_centers, np.full(len(coords_of_centers), g))
    hp.mollview(
        prob_map,
        coord=["Celestial", "Galactic"],
        min=0,
        max=1,
        title=f"Selection probability at G={g}",
    )
\end{lstlisting}

\end{appendix}

\end{document}